\documentclass[a4paper]{article}

\usepackage[english]{babel}
\usepackage[utf8x]{inputenc}
\usepackage[T1]{fontenc}

\usepackage[a4paper,top=3cm,bottom=2cm,left=3cm,right=3cm,marginparwidth=1.75cm]{geometry}

\usepackage{amsmath}
\usepackage[colorinlistoftodos]{todonotes}
\usepackage[colorlinks=true, allcolors=blue]{hyperref}
\usepackage{verbatim}
\usepackage{graphicx}
\usepackage{psfrag}
\usepackage{subfig}
\usepackage{array}
\usepackage{float}
\usepackage{authblk}
\usepackage{tikz}
\usetikzlibrary{arrows,positioning,calc}

\begin{document}

\title{Low Threshold Acquisition controller for Skipper CCDs}

\author[1]{Gustavo Cancelo}
\author[2]{Claudio Chavez} 
\author[3]{Fernando Chierchie}
\author[1]{Juan Estrada}
\author[3,1]{Guillermo Fernandez Moroni}
\author[3,4]{Eduardo Paolini} 
\author[5]{Miguel Sofo Haro}
\author[3]{Angel Soto}
\author[3,1]{Leandro Stefanazzi}
\author[1]{Javier Tiffenberg}
\author[1]{Ken Treptow}
\author[1]{Neal Wilcer}
\author[1]{Ted Zmuda}

\affil[1]{Fermi National Accelerator Laboratory, Batavia IL, United States.}
\affil[2]{Facultad de Ingenier\'{i}a - Universidad Nacional de Asunci\'{o}n, Paraguay, Asunci\'{o}n, Paraguay.}
\affil[3]{Instituto de Investigaciones en Ingenier\'{i}a El\'{e}ctrica ``Alfredo C. Desages''\\
Departamento de Ingenier\'{i}a El\'{e}ctrica y de Computadoras, Bahia Blanca, Argentina.}
\affil[4]{Comisi\'on de Investigaciones Cientif\'icas Prov. Buenos Aires (CICpBA), Argentina.}
\affil[5]{Centro At\'omico Bariloche and Instituto Balseiro, Comisi\'on Nacional de Energ\'ia At\'omica (CNEA), Universidad Nacional de Cuyo (UNCUYO), Rio Negro, Argentina.}

\maketitle

\begin{abstract}
The development of the Skipper Charge Coupled Devices (Skipper-CCDs) has been a major technological breakthrough for sensing very weak ionizing particles. The sensor allows to reach the ultimate sensitivity of silicon material as a charge signal sensor by unambiguous determination of the charge signal collected by each cell or pixel, even for single electron-hole pair ionization. Extensive use of the technology was limited by the lack of specific equipment to operate the sensor at the ultimate performance. In this work a simple, single-board Skipper-CCD controller is presented, aimed for the operation of the detector in high sensitivity scientific applications. The article describes the main components and functionality of the Low Threshold Acquisition (LTA) together with experimental results when connected to a Skipper-CCD sensor. Measurements show unprecedented deep sub-electron noise of 0.039\,$\rm e^-_{rms}/pix$ for 5000 samples.

\end{abstract}

\section{Introduction}

Charge-coupled devices (CCDs) have proven to be incredibly powerful for detecting photons and other particles that interact with the silicon atoms of the CCD. The number of applications in science and industry is still growing. For photon detection, CCD sensors rely on the photoelectric effect to absorb incident photons in a silicon substrate and generate electron-hole pairs \cite{janesick2001scientific}. Energetic photons (E>10 eV) produce multiple electron-hole pairs allowing an energy measurement. Instead lower energy photons may only generate one or very few electron-hole pairs. Massive particles can also create electron-hole pairs either by directly interacting with valence-band electrons or by scattering off silicon nuclei. For these reasons CCDs are not only used for optical applications but also for applications in nuclear physics and cosmology.

Most CCDs used in instruments such as cameras for astronomy and cosmology require high dynamic range. The CCD pixels collects the ionized charge generated by interacting photons during a certain  observation time. That signal may amount up to few hundred thousand electrons or holes (i.e. depending on the CCD type). Other applications in science and industry such as searches for dark matter and the coherent neutrino scattering, or neutron and quantum imaging, to name a few, are interested in faint signals, as small as 1 e-h pair. Henceforth, CCD dynamic range, pixel readout time and readout noise are application specific. However, independently of the application, precision measurements are limited by the readout noise of the CCD readout electronics. Other sources of measurement ambiguity, such as dark current are extremely low (e.g. $10^{-9}$e$^-$/pix/s) when the CCD is operating at low temperatures (e.g. 140K). Furthermore, charge transfer efficiency can be as good as $99.9999\%$ 
Recently, Tiffemberg et al \cite{Tiffenberg:2017aac} have successfully used a skipper-CCD to overcome the ~2~e$^-$ noise limit of standard CCDs approaching noise levels of 0.068~e$^-$ RMS. A Skipper-CCD has an extra charge storage that allows multiple non destructive measurements of the same pixel charge to be averaged in an uncorrelated fashion, breaking the $1/f$ dependence. At this noise level, the CCD becomes a e-h pair counter with an error probability of $\sim10^{-13}$. This Skipper-CCD designed by S. Holland at LBNL \cite{skipper_2012} is the first accurate single-electron counting silicon detector on a large-footprint. The Skipper-CCD also has linear gain which allows to readout signals of many thousand electrons with the same noise level. As pointed out in \cite{Tiffenberg:2017aac} that makes the Skipper-CCD the most sensitive and robust electromagnetic calorimeter that can operate at temperatures above that of liquid nitrogen. It also allows the Skipper-CCD to count individual optical and near-infrared photons. CCDs with sub-electron readout noise have a broad range of applications including particle physics (e.g., ultra-low-noise searches for dark matter and neutrinos) and astronomy (e.g., direct imaging and spectroscopy of exoplanets).The price we pay to achieve such an amazing noise level is in readout time. The 0.1e- RMS reported in \cite{Tiffenberg:2017aac} takes an averaging of 1200 reads of the same pixel charge using a pixel readout time of 10~$\mu s$ for a noise of 3.5~e$^-$/sample. That totals 12ms/pixel at 0.1~e$^-$ of noise. Even for applications that are not time critical, the readout time of a large device starts becoming a problem. In the following sections of this paper we will show that the noise-time performance of the Skipper-CCD can be improved by the CCD readout.
The reason for the improvement is simple. Readout electronics have been unable to attend the noise limit of the CCD video output. There is still room for a cleaner electronics that lowers the Correlated Double Sampling (CDS) noise integrated over a unit of time. A typical readout electronics also controls the charge sequencing out of the CCD and generating the bias voltages needed for operation. Both tasks introduce an excess noise into the CCD that adds to the intrinsic noise of the video signal path. The noise problem is analyzed in Section \ref{sec:noise theory}. 

Until now, the low threshold capabilities of these sensor were not able to be exploit in scientific or technological applications since the lack of a scientific grade equipment to operate it. Previous works with the sensor were reported using modified existing electronics \cite{skipper_2012,Tiffenberg:2017aac} in extremely well controlled experiments and without the possibility to scale or duplicate the system. The Low Threshold Acquisition readout electronics presented in this work, called LTA, is specifically designed to meet the requirements on this new scenario of growing demand of the technology. The group has large experience working with Skipper-CCD sensors and is currently leading cutting edge applications. Thanks to the compact design, the ease of fabrication, operation and noise performance, the LTA has already proven to be the solution for many groups working or willing to work with Skipper-CCDs. Among this applications we can cite: leading light dark matter searches currently being used in the SENSEI experiment; next generation of DAMIC experiment for dark matter searches; part of the R\&D program of the OSCURA project (10kg active silicon experiment for dark matter detection based on Skipper-CCDs); reactor neutrino detection; new spectrograph instruments; new single photon counting quantum imaging cameras; new program for high procession silicon property measurements (quenching factor, fano factor, ionization energy distribution, temperature dependence); cold neutron quantum measurements; construction of large Skipper-CCD testing labs; etc.

The paper is organized as follows: a description of the LTA system is presented in section \ref{sec:description}. A theoretical framework for the readout noise analysis in Skipper-CCD is given in section \ref{sec:noise theory}. Then, measurements of the performance of the controller disconnected from the CCD are presented in section \ref{sec:measurements LTA} and operating with a sensor in section \ref{sec:measurements with skipper}. Finally, some remarks and conclusions are presented in \ref{sec:conclusions}.
 

\section{Low Threshold Acquisition controller}
\label{sec:description}
\subsection{Hardware}

\begin{figure}
\begin{center}
\includegraphics[width=\textwidth]{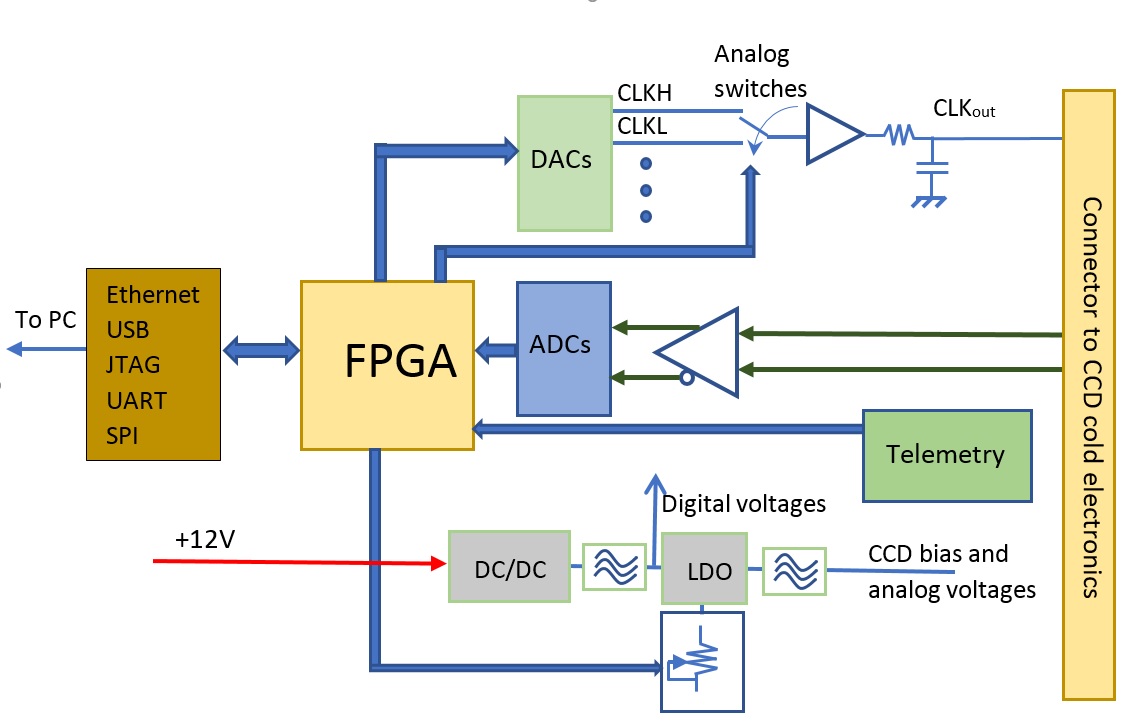}
\caption{Block diagram of the main hardware components in the LTA.}
\label{fig:block diagram}
\end{center}
\end{figure}

The LTA is a single PC board hosting 4 video channels for readout, plus CCD bias and control. It has been optimized to work with the p-channel, thick, high resistivity CCDs such as the ones at \cite{Holland:2003}. A block diagram of the LTA is given in Figure \ref{fig:block diagram}. Video channels are built using tightly-coupled low-noise differential amplifiers (LTC6363) and 18-bit, 15~Msps analog-to-digital converters (LTC2387) to capture the video signals, and digital correlated double sampling. A detailed characterization and description of the performance the former ADC applied for CCD readout can be found in \cite{haro2017low}. For a preamplifier gain of 4, to allow a large input dynamic range, the input noise floor of the LTA is -145dBFS/Hz.

The heart of the LTA is a Xilinx Artix XC7A200T FPGA. The FPGA is in charge of setting up the programmable bias voltages, managing the clock signals that move the pixel charge along the CCD array, video acquisition, telemetry and the shipment of data and status from the board to the PC. The fully-digital approach on the data path brings the advantage of digital signal processing techniques for noise reduction on the video channels. This feature of the LTA board has been previously explored in \cite{chierchiedetailed}, where an optimal filter with a better Signal to Noise Ratio (SNR) than a double slope integrator was presented. More about the FPGA firmware can be found in Section \ref{sec:firmware}.

\begin{figure}
\begin{center}
\includegraphics[width=\textwidth]{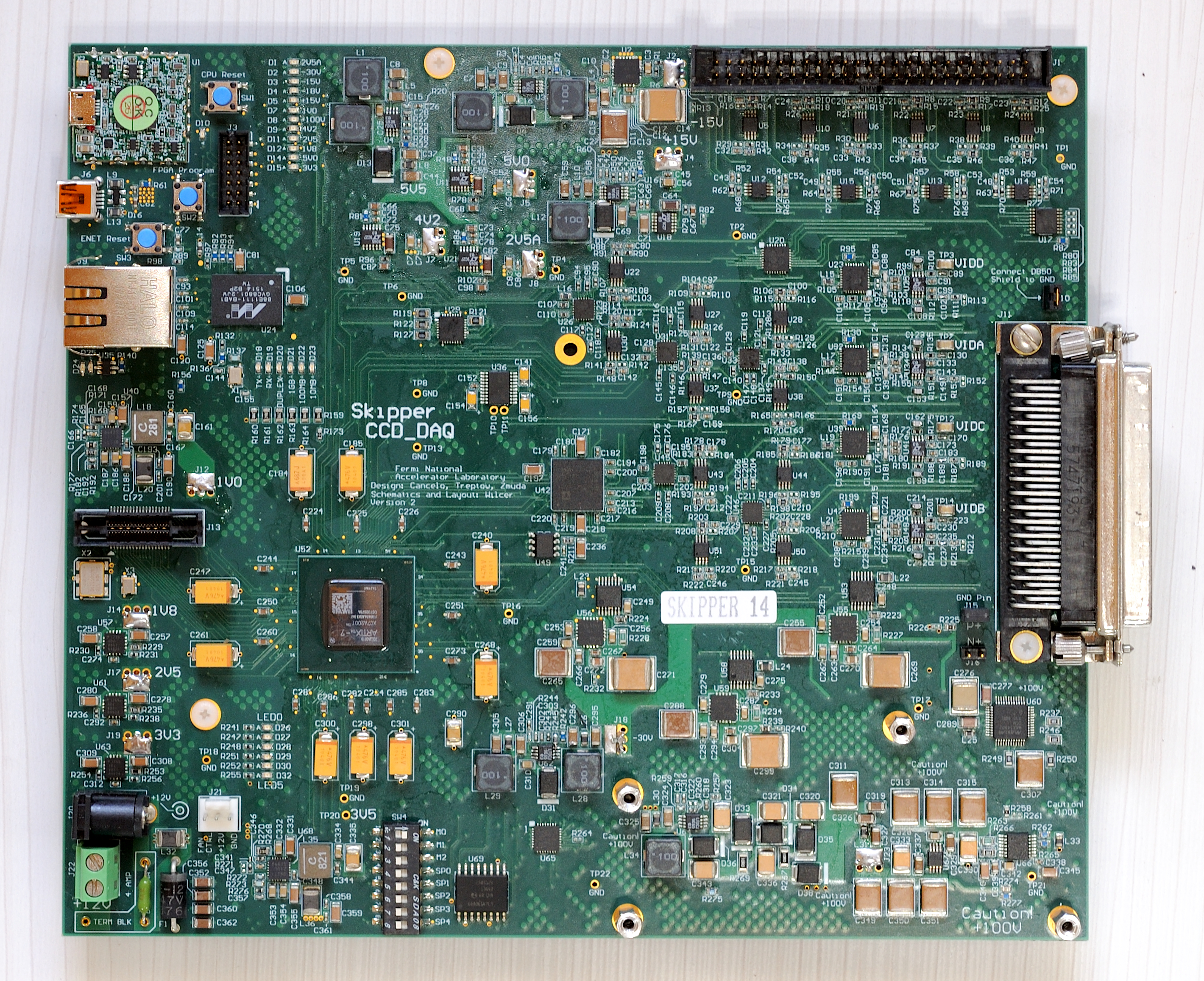}
        \caption{Low Threshold Acquisition electronics.}
        \label{fig:LTA picture}
\end{center}
\end{figure}

The CCD clocks are generated by a 40 channel DAC (AD5371), analog switches (ADG5234) and Op Amps (THS6072). The DAC is used as a multi-channel adjustable potentiometer. It allows setting clock values of up 20~V of dynamic range within a $\pm$~15~V. The voltage range is controllable by an internal offset. Each CCD clock uses two DAC outputs to define its low and high states. The analog switches are used to generate the clock signals switching between the low and high voltages of each clock. The Op Amps and associated R-C network provide the desired rise and fall time. Clock shaping and output noise are critical. As shown in Figure \ref{fig:block diagram} CCD clocking signals with voltage swings of up to 10 volts and dV/dt on the order 50V/usec are typical and induce crosstalk on the video lines. Although clocks remain stable during the video readout, some noise from clock feed-trough may still be present during CDS integration times. That is not White Gaussian Noise (WGN), so minimizing it at the generation point on the LTA board is of paramount importance. Coupling of 1ppm from a 10 V clock into the video has already a considerable negative impact on the noise.
Another critical part of the LTA is the power generation and management. The LTA is a 7 by 8 inches single board CCD readout and control (Figure \ref{fig:LTA picture}). A design of this kind requires many DC voltages to power digital and analog components. Furthermore, the CCD requires several DC bias voltages that must be as clean of noise as possible and have the flexibility of being adjusted for performance optimization.
The LTA only requires input power source of +12 Volts DC. Although we have not noticed any negative increase of noise using a standard AC/DC switching supply, it is important that the AC be kept clean \cite{haro2016measurement}. Nowadays AC lines are highly contaminated of high frequency switching noise. That noise couples to the output of either linear or switching AC/DC supplies. For extremely low noise measurements an AC noise filter are needed.

The DC switchers (LT3580) are operated at 2\,Mhz to keep switching noise away from the frequency band of interest for CCD readout. Since CCD bias are negative we used the ultra-low noise TPS7A33 with input and output filtering, achieving 100 $\mu$V of output noise RMS. The new series of ultra-low noise regulators are critical in minimizing the noise that couples to the video inputs on board and the noise that couples to the CCD and cables. This topic will be expanded in section \ref{sec:noise theory}. 

The CCD bias voltages are adjustable through AD5293 digitally controlled potentiometers. The power of other noise sensitive analog electronics such as the ADC and the video preamplifier are based on a combination of a LT3580 DC/DC switcher and the LT3045 ultra low noise linear regulator.

Special consideration has been taken with the CCD substrate bias Vsub which is controllable up to +100 Volts. The DC/DC converter uses an external voltage multiplier made with discrete components (i.e. diodes, capacitors and resistors). The output of the switcher is filtered by another ultra-low noise linear regulator TPS7A4001. Moreover, this hardware was designed to allows run the \textit{erase} procedure to the CCDs, in where Vsub is varied in a controlled way to reduce dark current and to remove ghost images \cite{Holland:2003}\cite{LTA8709274}. 

All power supplies and voltages default to strategic voltages at power up, and they are controlled and sequenced by the FPGA.
All clock and bias voltages are accessible through the telemetry system based on a AD7328 12-bit ADC and analog multiplexers. Also, a header connector on the board allows manual debugging of clock and bias signals through instrumentation such as an oscilloscope or logic state analyzer. The signals going to the header connector are buffered to prevent potential damage caused by the external instrumentation.
Besides the high-speed gigabit Ethernet interface, the board has USB, UART, SPI and JTAG interfaces for debugging purpose and FPGA programming. After debugging the firmware resides on an on-board Flash NVRAM and automatically boots up at power up.

The board also has dedicated hardware to synchronize many boards together in master-slave mode to build multiple CCD systems.

 \subsection{Firmware}
\label{sec:firmware}
\begin{figure}
\centering
\begin{tikzpicture}
[box/.style={align=center, draw,rounded corners,minimum width=1.5cm,draw=red!80,fill=red!50,line width=1.5pt},
arrow/.style={line width=2pt,draw=blue!50,latex'-latex'},
out box/.style={draw=green!50!black,dashed,line width=1pt,rounded corners},
box title/.style={fill=green!50,draw=green!80,line width=1.5pt,rounded corners}]

\node[box,minimum height=3cm,minimum width=2.5cm] (ublaze) {$\mu$Blaze\\processor};

\node[box,left=1.5 of ublaze,minimum height=1cm] (cds) {CDS};
\node[box,above=0.5cm of cds,minimum height=1cm] (buffer) {Smart\\buffer};
\node[box,below=0.5cm of cds,minimum height=1cm] (sync) {Sync};

\node[box,right=1.5of ublaze,minimum height=1cm] (packer) {Packer};
\node[box,above=0.5cm of packer,minimum height=1cm] (ethernet) {Eth};

\coordinate (a) at ($(ublaze.south)+(0,-1.5cm)$);
\coordinate (L1) at (cds.south |- a);
\coordinate (L2) at (packer.south |- a);

\node[box,anchor=north,minimum height=1cm] (telemetry) at (L1) {Tele};
\node[box,anchor=north,minimum height=1cm,minimum width=1.2cm] (clocks) at ($(L1)!0.35!(L2)$) {Clks};
\node[box,anchor=north,minimum height=1cm,minimum width=1.2cm] (bias) at ($(L1)!0.64!(L2)$) {Bias};
\node[box,anchor=north,minimum height=1cm] (sequencer) at (L2) {Seq};

\draw[arrow]
	(telemetry.north) --
	++(0,0.7) coordinate (tb) --
	(tb -| sequencer.north) --
	(sequencer.north);
	
\draw[arrow,latex'-]
	(clocks.north) --
	(clocks.north |- tb);
	
\draw[arrow,latex'-]
	(bias.north) --
	(bias.north |- tb);

\draw[arrow,latex'-]
	(ublaze.south) --
	(ublaze.south |- tb);
	
\draw[arrow,latex'-]
	($(buffer.south east)!0.3!(buffer.north east)$)  --
	++(0.7,0) coordinate (sb) --
	(sb |- tb);	
	
\draw[arrow,latex'-]
	($(cds.south east)!0.3!(cds.north east)$)  coordinate (temp) --
	(temp -| sb);

\draw[arrow,latex'-]
	($(ethernet.south west)!0.3!(ethernet.north west)$)  --
	++(-0.7,0) coordinate (eb) --
	(eb |- tb);	

\draw[arrow,latex'-]
	($(packer.south west)!0.3!(packer.north west)$)  coordinate (temp) --
	(temp -| eb);
	
\draw[arrow,latex'-]
	($(sync.south east)!0.5!(sync.north east)$)  coordinate (temp) --
	(temp -| sb);
	
\coordinate (tb p) at ($(tb)+(0,0.25)$);	
\coordinate (sb) at ($(buffer.south east)!0.6!(buffer.north east)+(1,0)$);
\coordinate (eb) at ($(sb -| ethernet.west)+(-1,0)$);
\draw[arrow,yellow!70!black]
	(buffer.east |- sb) --
	(sb) --
	(sb |- tb p) --
	(tb p -| eb) coordinate (temp) --
	(temp |- eb) --
	(eb -| ethernet.west);

\draw[arrow,yellow!70!black,latex'-]
	($(cds.south east)!0.6!(cds.north east)$) coordinate (temp) --
	(temp -| sb);
	
\draw[arrow,yellow!70!black,latex'-]
	($(packer.south west)!0.6!(packer.north west)$) coordinate (temp) --
	(temp -| eb);	

\draw[out box]
	($(telemetry.south west)+(-0.4cm,-0.4cm)$) coordinate (temp 1)
	rectangle
	($(ethernet.north east)+(0.4cm,0.4cm)$) coordinate (temp 2);


\draw[arrow,latex'-,line width=1pt,yellow!70!black]
	($(buffer.north west)!0.2!(buffer.south west)$) --
	++(-1,0) coordinate (a);
	
\draw[arrow,latex'-,line width=1pt,yellow!70!black]
	($(buffer.north west)!0.4!(buffer.south west)$) coordinate (temp)--
	(temp -| a) coordinate (b);

\draw[arrow,latex'-,line width=1pt,yellow!70!black]
	($(buffer.north west)!0.6!(buffer.south west)$) coordinate (temp)--
	(temp -| a) coordinate (c);

\draw[arrow,latex'-,line width=1pt,yellow!70!black]
	($(buffer.north west)!0.8!(buffer.south west)$) coordinate (temp)--
	(temp -| a) coordinate (d);

\draw[arrow,latex'-,line width=1pt,yellow!70!black]
	($(cds.north west)!0.2!(cds.south west)$) coordinate (temp)--
	++(-0.55,0) coordinate (temp) --
	(temp |- a) coordinate (temp);
\draw[arrow,yellow!70!black] (temp) circle (0.5pt);
	
\draw[arrow,latex'-,line width=1pt,yellow!70!black]
	($(cds.north west)!0.4!(cds.south west)$) coordinate (temp)--
	++(-0.65,0) coordinate (temp) --
	(temp |- b) coordinate (temp);
\draw[arrow,yellow!70!black] (temp) circle (0.5pt);

\draw[arrow,latex'-,line width=1pt,yellow!70!black]
	($(cds.north west)!0.6!(cds.south west)$) coordinate (temp)--
	++(-0.75,0) coordinate (temp) --
	(temp |- c) coordinate (temp);
\draw[arrow,yellow!70!black] (temp) circle (0.5pt);

\draw[arrow,latex'-,line width=1pt,yellow!70!black]
	($(cds.north west)!0.8!(cds.south west)$) coordinate (temp)--
	++(-0.85,0) coordinate (temp) --
	(temp |- d) coordinate (temp);
\draw[arrow,yellow!70!black] (temp) circle (0.5pt);

\draw[arrow,latex'-]
	(telemetry.south) --
	++(0,-1);
	
\draw[arrow,-latex']
	(clocks.south) --
	++(0,-1);

\draw[arrow,-latex']
	(bias.south) --
	++(0,-1);
	
\draw[arrow,-latex']
	(sequencer.south) --
	++(0,-1);
https://www.overleaf.com/project/5e750d164f07e60001a3849e
\draw[arrow,yellow!70!black]
	(ethernet.east) --
	++(1,0);

\draw[arrow,latex'-,line width=1pt,yellow!70!black]
	($(sync.north west)!0.2!(sync.south west)$) --
	++(-1,0) coordinate (a);
	
\draw[arrow,-latex',line width=1pt,yellow!70!black]
	($(sync.north west)!0.4!(sync.south west)$) coordinate (temp)--
	(temp -| a) coordinate (b);

\draw[arrow,latex'-,line width=1pt,yellow!70!black]
	($(sync.north west)!0.6!(sync.south west)$) coordinate (temp)--
	(temp -| a) coordinate (c);

\draw[arrow,-latex',line width=1pt,yellow!70!black]
	($(sync.north west)!0.8!(sync.south west)$) coordinate (temp)--
	(temp -| a) coordinate (d);
	
\end{tikzpicture}
\caption[]{Detail of the FPGA firmware with low-speed control AXI bus (\tikz[baseline=-0.5ex] \draw [line width=2pt,blue!50] (0,0) -- ++(1.5ex,0);) and high throughput data paths (\tikz[baseline=-0.5ex] \draw [line width=2pt,yellow!70!black] (0,0) -- ++(1.5ex,0);).}
\label{Fig:FpgaFirmware}
\end{figure}
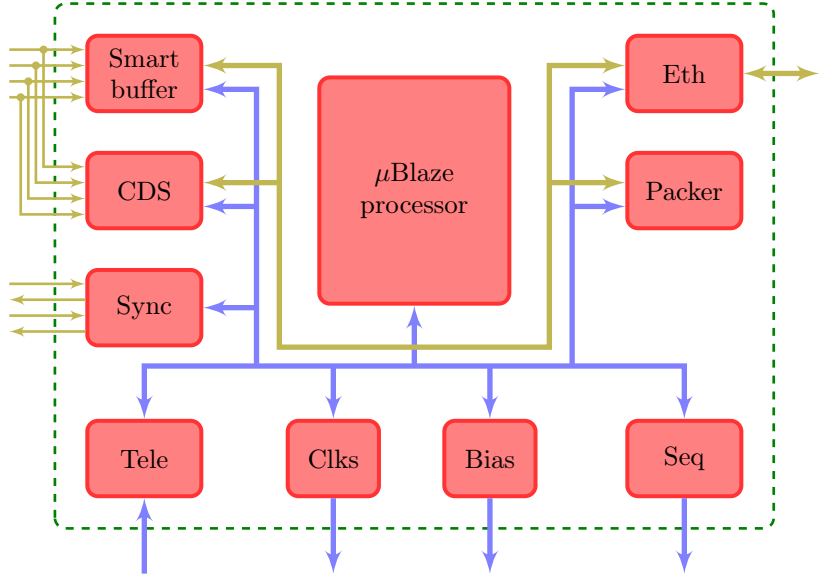

The user interacts with the board through a single Ethernet port, which allows sending and receiving commands as well as data. Serial interface is reserved for debugging purposes. A block diagram of the main units and control flow of the firmware on the FPGA is shown in Fig. \ref{Fig:FpgaFirmware}. The initial setup and the slow speed control of all the blocks are driven by a soft-core $\mu$Blaze processor. The software is a stand-alone, lightweight application developed in C-language, requiring not more than 256~kByes of memory, which allows the use of internal FPGA memory.

Precise timing is required to control the clock signals of the CCD. To this end, the sequencer module (Seq) is implemented in hardware to ensure precise synchronization of clock signals and ADC samples processing during the CCD readout. In normal operation, ADC samples are fed into the Digital Correlated Double Sampling (CDS) block, which computes the average of the samples. The integration window is controlled by the sequencer. 

The user can also send ADC data into the Smart Buffer block for raw data transmission. 
It should be clear that it is not possible to store a full image on the FPGA's internal memory for further analysis, and due to the speed of the raw data stream (15 MSPS with 18-bit), it is not possible to transfer these data on real time. As a result of these limitations, the Smart buffer stores the raw data on a controlled manner. This block was developed to behave like an integrated oscilloscope, which means the user can configure it to capture one, two or the four raw video channels. Single or continue capture is also supported. As an extra feature, signals from the sequencer are routed to the buffer and are used like external triggering, allowing to store data samples at very precise moments. Data stored by the Smart Buffer is kept until the user decides to transfer it using the Ethernet interface using a lower speed which can also be configured on the fly. This oscilloscope mode is particularly important when optimizing readout speed and noise performance or detecting problems in the video chain. 

In normal operation, the pixel information coming out of the CDS block is wrapped with extra control information in the Packer module and sent to the Eth module for external transmission. Clks and Bias blocks control de configuration of clocks and bias voltages, and the Tele module is in charge of the telemetry information collection from clocks and biases voltages. 
The LTA also allows for easy scaling to any number of CCDs operated in parallel. For this purpose the boards are connected together and configured in slave mode. Only one of them runs in master mode providing synchronization of the CCD clock sequence for the entire group. A synchronized readout avoids extra noise from cross talk noise from clocks transitions.  

JTAG port connectivity is added for easy programming and debugging. Non-volatile 512\,MB flash memory provides location to store the FPGA bit-stream for permanent, non-JTAG functioning.

 \subsection{Software and DAQ}

LTA software is written in C++ with open sources libraries for IP communications. It is based on a client-server model where a local daemon provides services for commands, telemetry and data acquisition.  
Users interact with the software using a request-response method through the terminal to perform  LTA board configuration, readout and  telemetry requests and also sequencer uploading.

Data and commands are transmitted entirely using UDP over IP through the computer giga-etherner port and the software relies on the hard-drive as buffer to avoid unnecessary RAM usage. At the end of data acquisition the software performs a coherency check and delivers final images in FITS and other formats.

The code can be compiled in several Linux distributions such as Ubuntu, Scientific Linux, Rasp-bian, Ubuntu over Windows WSL,  Open Suse over Windows WSL. With few modifications it can run over other Linux distributions.

 \section{Theoretical noise framework}
 \label{sec:noise theory}
 \subsection{Noise sources}
\begin{figure}
\begin{center}
\includegraphics[width=\textwidth]{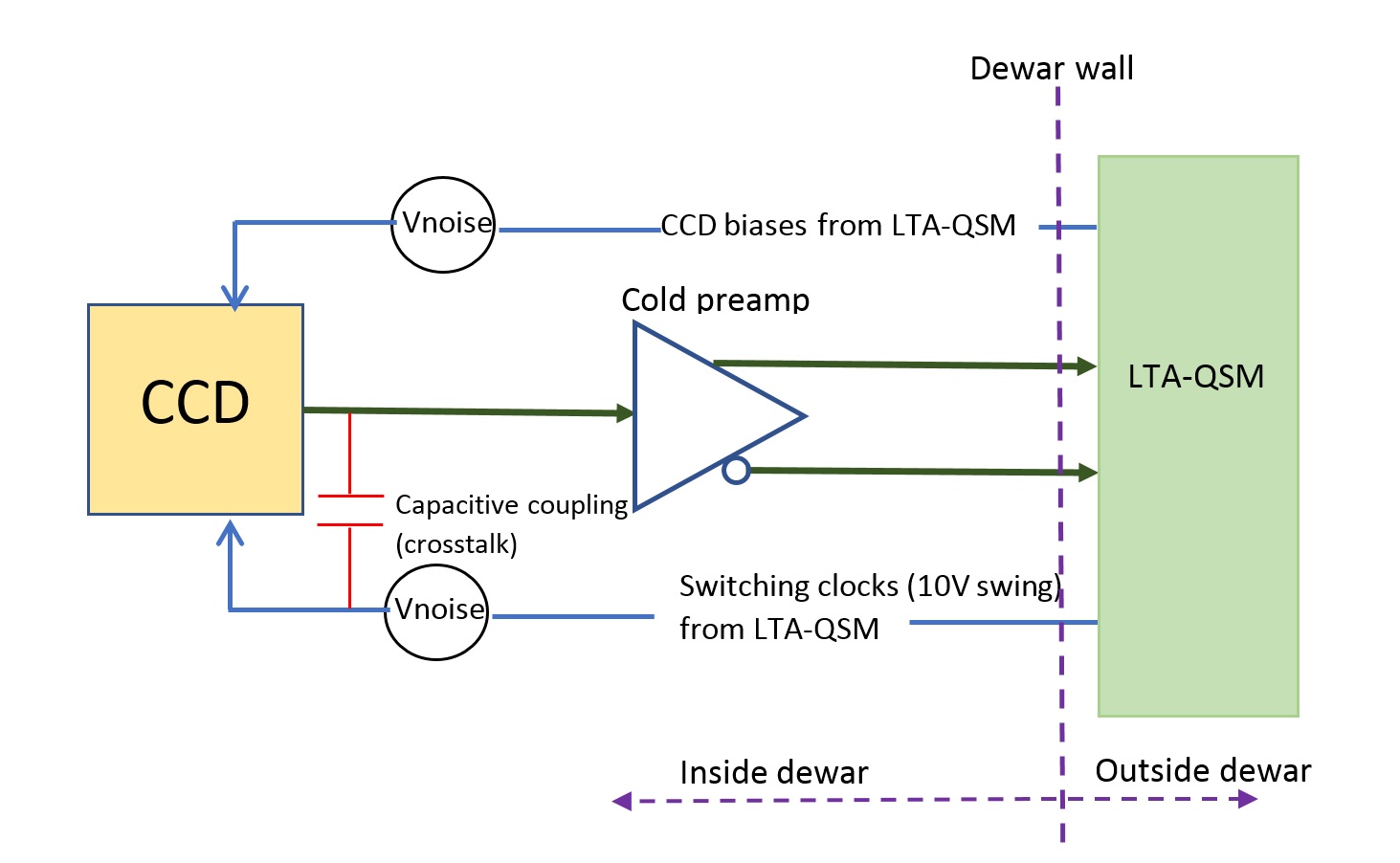}
        \caption{Noise sources block diagram.}
        \label{fig:noise sources}
\end{center}
\end{figure}

 The CCD sensor and its associated control and readout electronics constitute a tightly coupled system. The total readout noise is a combination of all noise sources along the readout chain. That chain starts with the internal CCD readout MOS-FET transistor, continues with an interface electronics for preamplification and finishes with the data acquisition and video signal processing. Noise from control and bias signals  will couple to the CCD internal readout and also along the cabling (e.g. CCD package, flex cables, etc.).  Since all control and bias signals are highly coupled to the video signal they are potential noise sources. Figure \ref{fig:noise sources} summarizes the main noise sources to be considered. Minimizing the control and bias signal noise and the additive noise from external amplification stages is extremely important and allows to get the lowest uncertainty limit given by the built in output transistor of the sensor.
 
 \subsection{Correlated Double Sampling}
 
Since the CCD pixel array is readout serially, through one or more output amplifiers, the amplifier’s noise is added to the charge of each pixel \cite{janesick2001scientific}. The readout method to achieve lower noise is integration. If we assume that most of the noise energy is white and Gaussian, we obtain a noise reduction proportional to the square root of the integration time. Furthermore, as most integrators require a reset, a pixel baseline is needed which comes with a time penalty of equal integration time. That readout method, widely called Correlated Double Sampling (CDS) \cite{janesick2001scientific} is described by the equation
 
\begin{alignat}{1}
P_i = \frac{1}{t_i} \left( \int_{t_i+\tau}^{2t_i+\tau} x(t) dt - \int_{0}^{t_i} x(t) dt\right),\label{eq:2}
\end{alignat}
where $P_i$ is the final pixel value, $\tau$ is the time spent to transfer the charge to the sensing node after the reset, $A$ is an arbitrary gain of the video chain, $t_i$ is the integration time. The first integral on the left side represents the integration over the charge and the second is over the pedestal (reference value). The integrals are normalized by the integration time $t_i$. The CDS can be seen as a filter with a frequency domain transfer function
\begin{equation}
|H_{CDS}(f)|=\frac{2A}{\pi t_i f}\sin ^{2}\left(\pi t_i f \right).
\label{eq:CDS magnitud}
\end{equation}

The output noise power spectrum ($S_{yy}$) is equal to the input noise power spectrum ($S_{xx}$) times the square of equation \ref{eq:CDS magnitud}
\begin{alignat}{1}
S_{yy} = S_{xx} \left|H(f)\right|^2 \label{eq:4}\\
\end{alignat}

\begin{figure}
 	\centering
 	\includegraphics*[width=0.75\columnwidth]{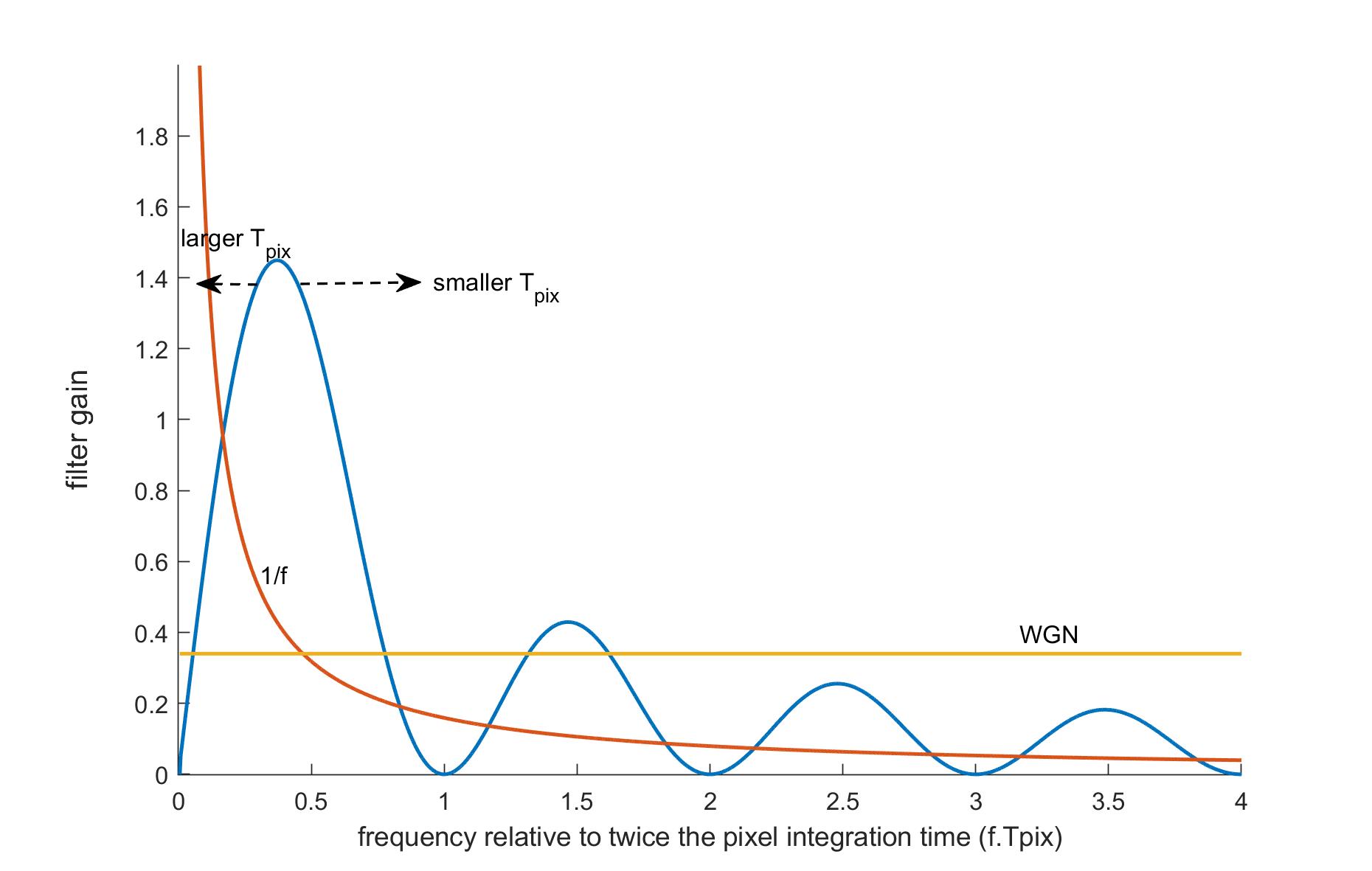}
 	\caption{CDS transfer function example with the characteristic noise spectra in the CCD video signal.}
 	\label{fig:CDS simulation}
 \end{figure}

Figure \ref{fig:CDS simulation} shows the magnitude of the CDS filter $\left|H_{CDS}(f)\right|$ as a function of the normalized frequency $t_if $. The filter band-pass is maximum at a frequency of $t_i$. Since the numerator is bounded and periodic and the denominator is proportional to $f$, the filter attenuates high frequencies. At low frequencies the filter response decays towards zero gain, again filtering the WGN flat spectrum. However, using Taylor expansion and the L'Hopital's approximation it can be shown that as $f$ approaches zero the integral of the $1/f$ noise is constant and independent of the integration time.

CDS is optimal only for WGN because the standard deviation approaches 0 as the integration time goes to $\infty$. Non WGN noise such as $1/f$ low-frequency readout noise and other non WG system level readout noise has remained a fundamental limitation for precision single-photon and single-electron counting in CCD based systems. In conventional scientific CCDs, low-frequency readout noise results in root-mean-squared (rms) variations in the measured charge per pixel at the level of $\sim$\,2\,$\rm e^-_{rms}/pixel$ \cite{Holland:2003}\cite{CONNIE_2019}\cite{Aguilar-Arevalo:2016ndq}\cite{haro2016measurement}\cite{haro2016taking}. For longer integration times the noise plateaus or even starts raising.

Similar results are obtained assuming a Digital Correlated Double Sampling (CDS) as is the case for the LTA. In this case the integration time should be seen as the number of ADC samples ($n_i$) being average multiplied by the ADC sampling time ($T_S=1/F_S$), $t_i = n_iT_S$.

\subsection{Skipping readout to overcome 1/f barrier}

\begin{figure}
 	\centering
 	\includegraphics*[width=0.75\columnwidth]{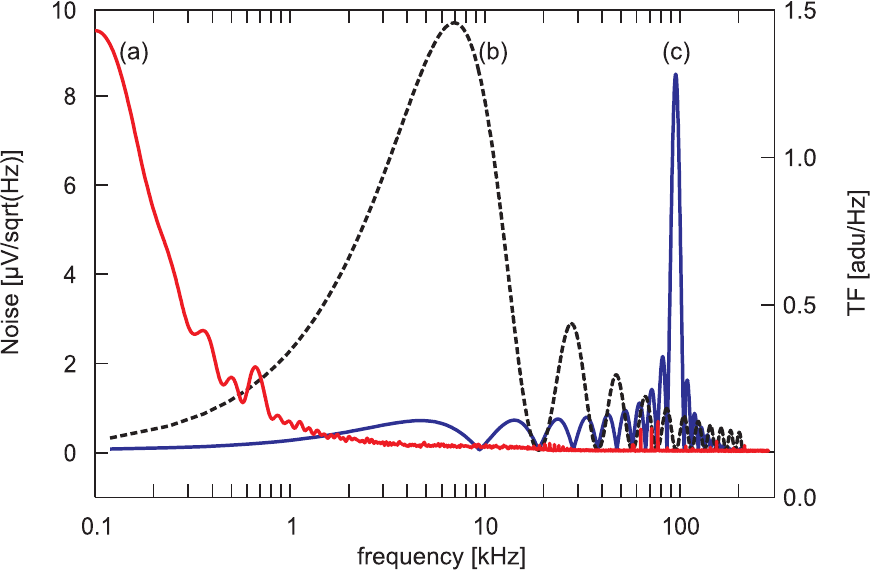}
 	\caption{Left axis Depletion MOS-FET noise PSD at a temperature of 273 K (a). Right axis frequency response of CDS readout system for $t_i$ =55~$\mu$s (b)and of the Skipper-CCD readout system for $t_i=5.5\mu$ and N=10(c). Figure taken from \cite{skipper_2012}.}
 	\label{fig:SkipperCCDRFN}
 \end{figure}

A Skipper-CCD has an extra charge storage that allows multiple non destructive measurements of the same pixel charge to be averaged in an uncorrelated fashion, breaking the $1/f$ dependence. In this case, the final pixel value is %
\begin{alignat}{1}
P_i = \frac{1}{N}\sum^{N}_{j=1}\frac{1}{t_i} \left( \int_{t_i+\tau+t_j}^{2t_i+\tau+t_j} x(t) dt - \int_{t_j}^{t_i+t_j} x(t) dt\right),\label{eq:2}
\end{alignat}
 where $t_j$ is the initial time of each measurement of the pixel charge. The corresponding transfer function magnitude (from \cite{skipper_2012}) is

 \begin{equation}
|H_{SKP}(f)|=\frac{2A}{\pi N t_i f}\sin ^{2}\left(\pi t_i f \right)\left\vert \frac{\sin (\pi N t_ f)}{\sin(\pi t_i f)}\right\vert .
 \label{eq:HSKIP}
 \end{equation}
The main advantage of the Skipper-CCD readout system is revealed when its frequency response is compared to the standard CDS frequency response for the same total integration time ($t_T$), i.e., using the total amount of time for all the samples in the skipper readout $t_T = Nt_i$. In Fig. \ref{fig:SkipperCCDRFN} the frequency response $|H_{CDS}(f)|$ of the CDS readout system (curve (b)) is plotted for $t_i = 55\mu$ s, together with the frequency response of the Skipper-CCD readout system for $t_i=5.5\mu$s and N = 10 (curve (c)), using a logarithmic frequency axis. The noise PSD measured for a CCD output amplifier at 273 K is also included (curve (a)) from \cite{skipper_2012}. Both systems achieve the same level of white noise reduction because the total integration time $t_T$ of the video signal is the same in both cases. However, the gain of Skipper-CCD readout system is much lower than CDS readout system at low frequencies, allowing a readout noise reduction in presence of low frequency noise. This reduction can be increased by merely augmenting the number N of averaged samples. In other words, low frequency noise filtering is achieved and the LF limitation is removed, at the only expense of increasing readout time. This expectations has been extensively proven in \cite{skipper_2012, Tiffenberg:2017aac}.

Infinite convention of integration times ($t_i$) and number of measurements per pixel ($N$) can be used to obtain a desired overall readout noise of the system. Their optimum values should be optimized based on the application. For example in \cite{skipper_2012}, the authors show how the optimization is performed to minimize the effect of both readout noise and dark current contributions at the same time. For the purpose of this article, the performance of the LTA will be shown for several scenarios of $t_i$ and $N$.

\subsection{Conversion factor for equivalent noise charge (ENC) measurements}

Gain calibration of the video signal chain plays a significant role in the normalization of the noise quantities referred to the charge signal being measured. Typical gain calibration procedures using X-rays or the photon transfer curves \cite{janesick2001scientific} rely on large charge packets values per pixel which relay on a very well known mean ionization energy for X-rays at a given temperature, and a uniform exposure for the photon transfer curve. To avoid any systematic error from these procedures, gain calibration is performed exploring the single charge counting capability of the Skipper-CCD. Using deep subelectron readout noise levels, the pixel value can be fully discretized to an integer number of collected carriers. This enables an absolute and very precise calibration of the gain of the video chain as the difference between the pixel values of charge packets differing in one collected electron. This allows for a precise calibration in the entire dynamic range of the system. In section \ref{sec:measurements with skipper}, we exploit this capability to calibrate our system using pixels with zero and one colleected electrons.

\section{Intrinsic noise performance of the LTA}
\label{sec:measurements LTA}

This section looks at the LTA electronics disconnected from the CCD and investigates:
 description and noise characteristics of the interface electronics used to connect the LTA to the sensor; the noise at the LTA input given by the preamplifier, ADC and all coupled noise from the rest of the board;
 the noise generated by the LTA control and bias lines.

\subsection{Noise performance of the interface electronics}

\begin{figure}
\begin{center}
\includegraphics[width=\textwidth]{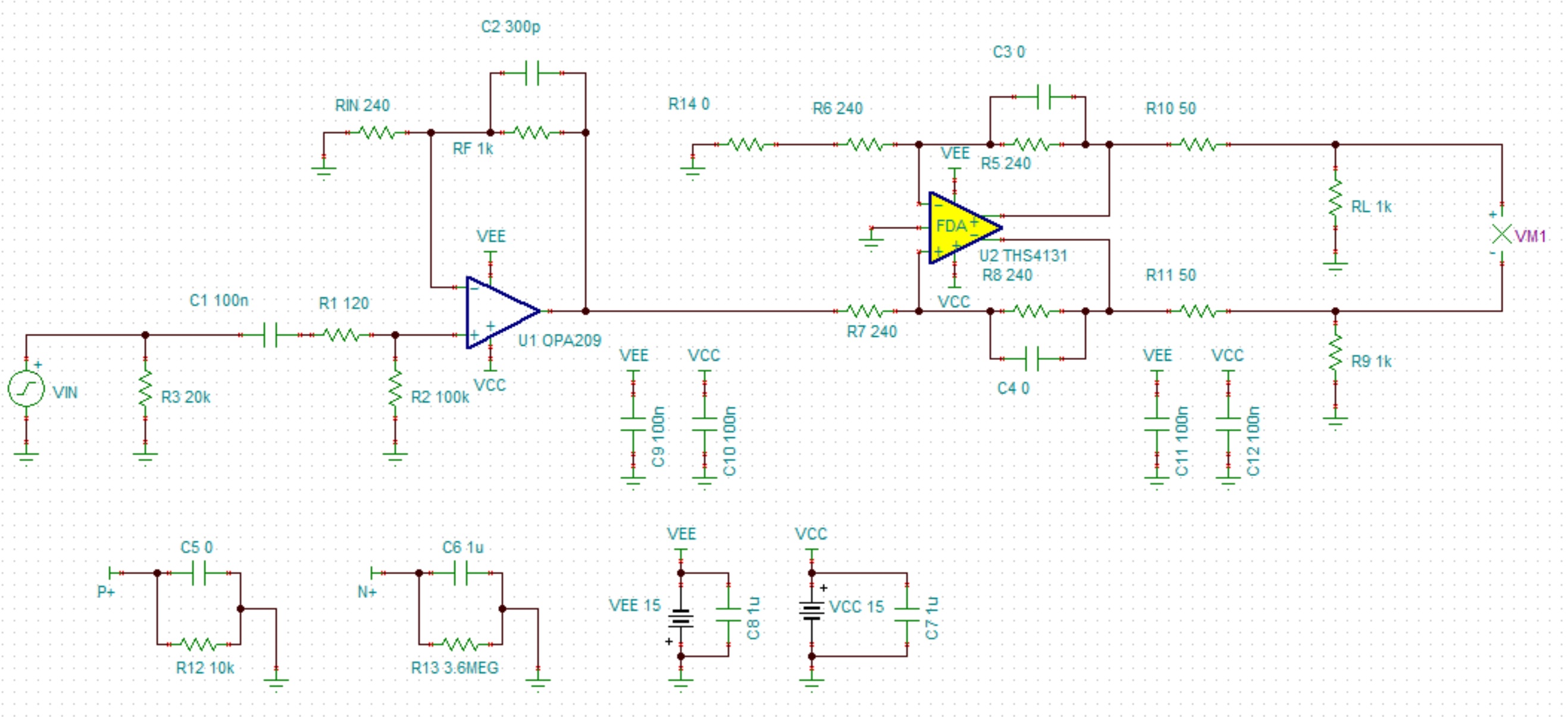}
        \caption{Interface electronics schematic.}
        \label{fig:cold electronics}
\end{center}
\end{figure}

\begin{figure}
\begin{center}
\includegraphics[width=\textwidth]{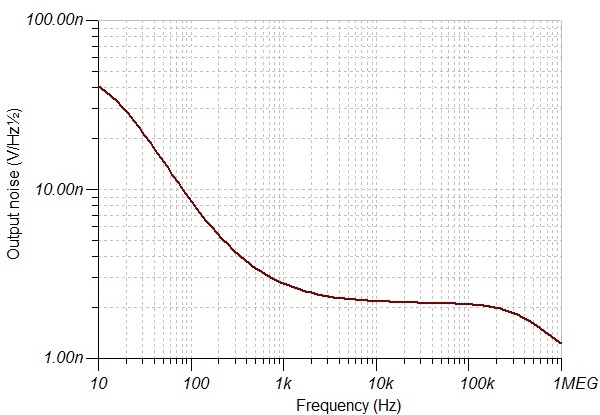}
        \caption{Interface electronics noise spectrum.}
        \label{fig:transition electronics noise spectrum}
\end{center}
\end{figure}

\begin{figure}
\begin{center}
\includegraphics[width=\textwidth]{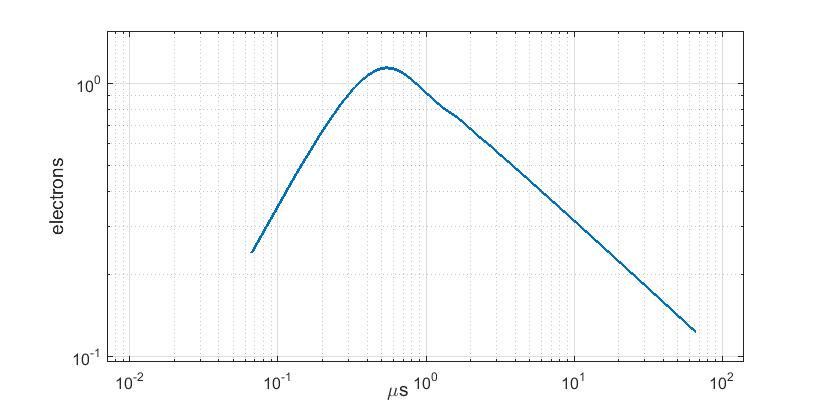}
        \caption{Interface preamplifier noise contribution after CDS.}
        \label{fig:cold electronics noise after CDS}
\end{center}
\end{figure}

The typical interface electronics used to connect the LTA to the sensor is a two stage preamplifier (Figure \ref{fig:cold electronics}). The first stage must decouple the CCD DC bias while keeping a zero close to DC to avoid signal and pedestal drooping. At the same time the noise added by that stage must be minimized. That is accomplished with the OPA209 in non inverting configuration. The nominal gain of the stage is 5 but lower or larger values are possible to adjust the dynamic range. A high frequency pole is introduced on the feedback loop to minimize the output noise. The output noise spectrum referred to the input is depicted in Figure \ref{fig:transition electronics noise spectrum}. At 140~K degrees the $1/f$ noise contribution at 100 Hz is an order of magnitude smaller than the CCD's video transistor. The WGN floor is $\sim 2~nV /\sqrt{Hz}$ compared to $\sim 20~nV /\sqrt{Hz}$ of the CCD JFET. The noise spectrum shows a knee at about 1 MHz given by the high frequency pole to minimize the total noise. The total noise of the stage for 1~MHz bandwidth is $1.7~\mu $. A simulation of the noise contribution of the cold electronics noise in the CDS filtering is shown in Figure \ref{fig:cold electronics noise after CDS}. The maximum noise is achieved at $0.55~\mu s$ of integration time and it decays as $\frac{1}{\sqrt{T}}$ for larger integration times. The plot has been calibrated in electrons using a measured $2.5~\mu$V/e$^-$ gain. At $T=1~\mu$s the noise is a fraction of one electron. 

The purpose of the second stage is to convert the signal to differential output to reject common mode noise coupling on long cables running from the dewar to the LTA electronics. Those cables are not needed for the current setup but they will be needed for multi-CCD application experiments such as SENSEI. SENSEI will have 50 CCDs and a total of 200 video channels, requiring 50 LTA boards.

\subsection{Noise performance of the LTA}
\begin{figure}
\begin{center}
\includegraphics[width=\textwidth]{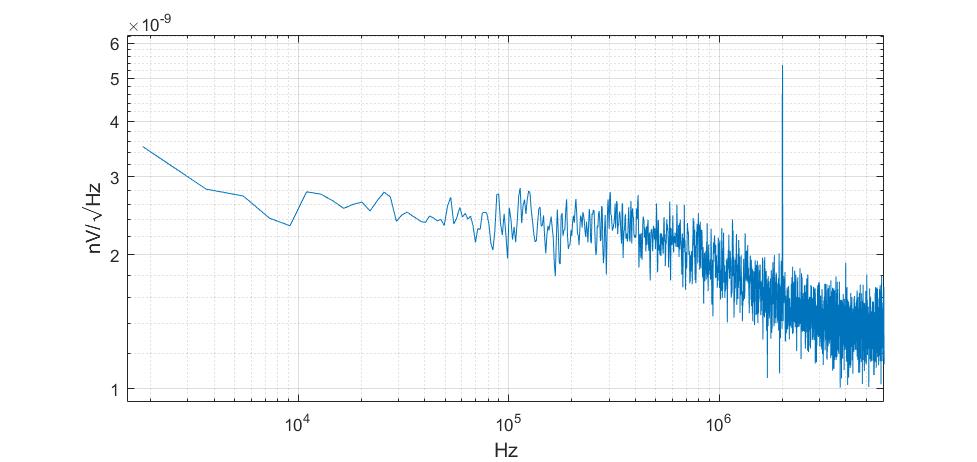}
        \caption{50-ohm-terminated noise spectrum.}
        \label{fig:wls1}

\end{center}
\end{figure}

\begin{figure}
\begin{center}
\includegraphics[width=\textwidth]{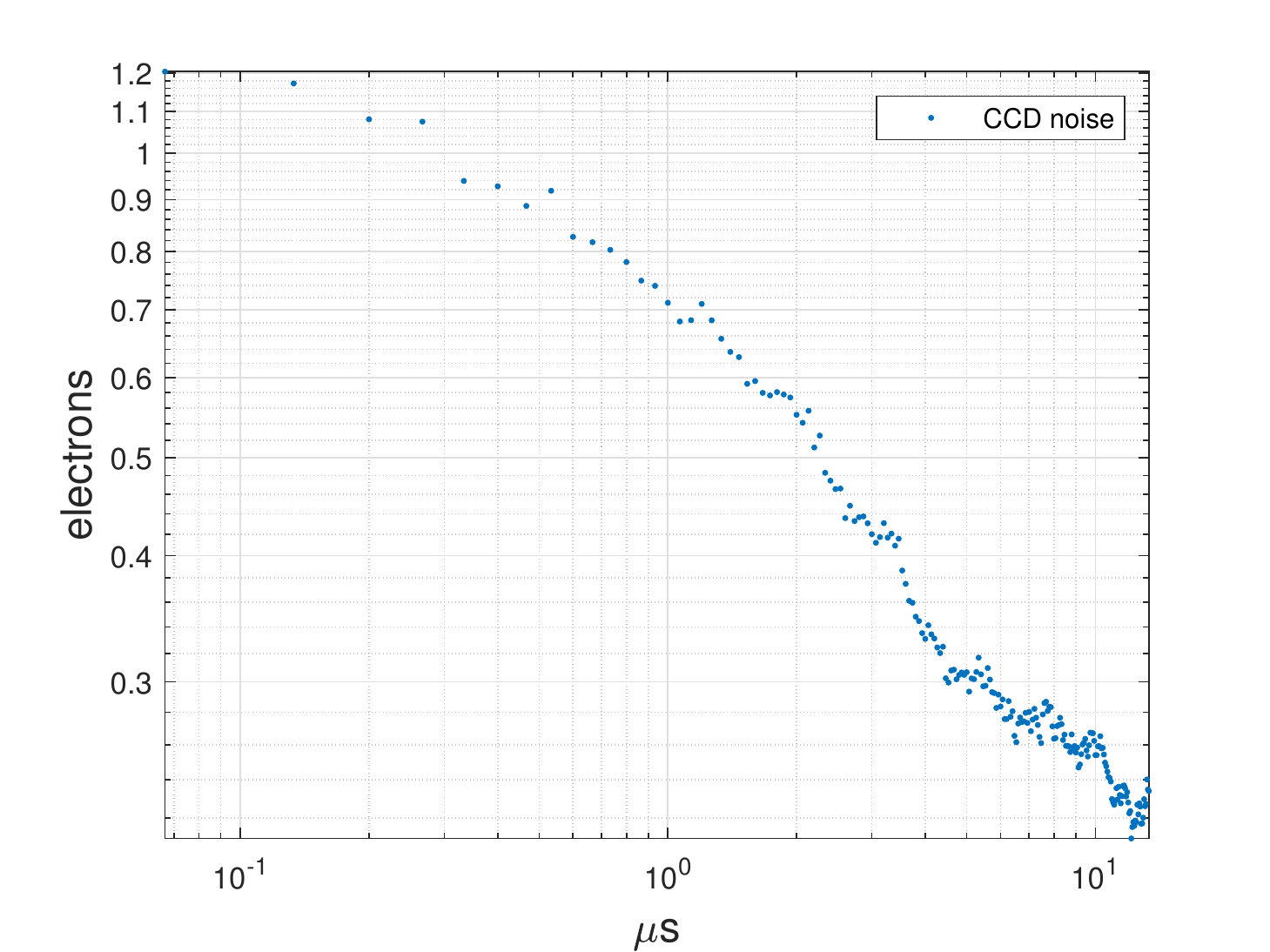}
   \caption{Output noise level at the output of the CDS using the 50-ohm-terminated noise spectrum.}
    \label{fig:wls2}
\end{center}
\end{figure}

A measurement of the noise of one of the the four LTA video inputs when terminated by a $50 \Omega$ load is shown in Figure \ref{fig:wls1} and \ref{fig:wls2}. That load is equivalent to the output impedance of a low noise preamplifier located near the CCD (e.g. the CCD package or flex cable). The rest of the signals, bias and clocks are loaded with R-C networks to emulate a CCD. Figure \ref{fig:wls1} shows the noise spectrum when the clocks are held constant at nominal readout levels. Figure \ref{fig:wls2} shows the CDS noise as a function of the integration time ($t_i$) when all the clocks switch as in a normal readout operation. The spectrum is 1/f  dominated up to 10 kHz and is WG until it finds the filter's pole located close to 1MHz.

 The RMS of the noise for 1MHz bandwidth is $20~\mu$V at the input of the LTA and represents 2.5 ADC counts when the internal preamplifier gain is 4. We typically use a transition preamplifier stage between the CCD and the LTA with a gain between 5 and 10. For a preamplifier gain of 10 the LTA noise referred to the input of the transition preamplifier (i.e. the CCD video output) is $2~\mu V$ RMS. This RMS noise is smaller than 1 electron and when integrated by the CDS algorithm becomes negligible. Figure \ref{fig:wls2} shows the contribution of the LTA noise versus CDS integration time referred at the input of the interface preamplifier.

\begin{figure}
\begin{center}
\includegraphics[width=\textwidth]{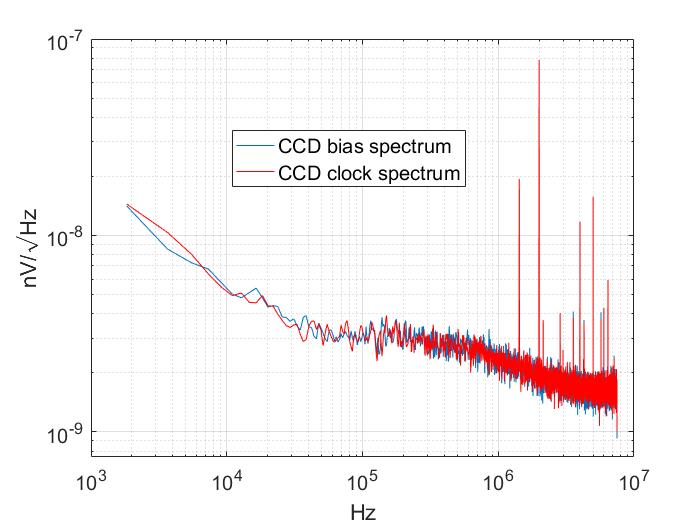}
        \caption{Noise spectrum of the analog signals controlling the sensor.}
        \label{fig:bias and clock noise spectrum}
\end{center}
\end{figure}

The CCD control and bias are sources of noise (Figure \ref{fig:noise sources}). That noise capacitively couples to the video through the CCD and in the CCD package and cables. High $\frac{dV}{dt}$ noise is bad particularly during the readout integration times. Figure \ref{fig:bias and clock noise spectrum} shows the noise spectrum of one of the LTA bias signals and one of the clock signals. Since all the bias use the same DC/DC converters, filters and LDOs, that spectrum is representative for all biases. The same happens with the other clock signals, the spectrum shown is representative of the entire group. 
Both spectra show a flat WG noise floor of $\sim 3~nV/\sqrt{Hz}$. The $1/f$ noise meets the WGN at $\sim 30KHz$ and its power is similar to the $1/f$ noise power of the internal video transistor (i.e. source follower of the CCD). However, only a small fraction of the bias and clock noise will couple to the video lines due to the CCD readout circuit CMRR (Common Mode Rejection Ratio). 
The coupling of noise to the CCD package and cables is mostly capacitive. Henceforth, it is more important at higher frequencies and larger $\frac{dV}{dt}$. The spectra of Figure \ref{fig:bias and clock noise spectrum} show tall lines of up to 20 dB higher than the noise floor. These lines correspond to the DC/DC converter switching frequencies 2MHz and harmonics. There are also smaller lines corresponding to internal oscillators for the digital electronics, mainly the FPGA. These lines are partially rejected by the 1MHz analog filters at the cold electronics preamplifier and at the LTA preamplifier, but most importantly by the CDS. For a typical integration time of $10~\mu s$ the CDS attenuates a 2~MHz spike by more than 40~dB, making noise contribution above 2 MHz negligible.

To reject pickup noise along the cables and noise coupled from the dewar the interface electronics and LTA input preamplifiers are differential. The CMRR of each stage is above 60~dB for frequencies above 1~MHz.



\section{Noise performance of the LTA with a Skipper-CCD}
\label{sec:measurements with skipper}

In this section we discuss the performance of the LTA operating a Skipper-CCD. 

\subsection{The readout setup}
\begin{figure}
\begin{center}
\includegraphics[width=\textwidth]{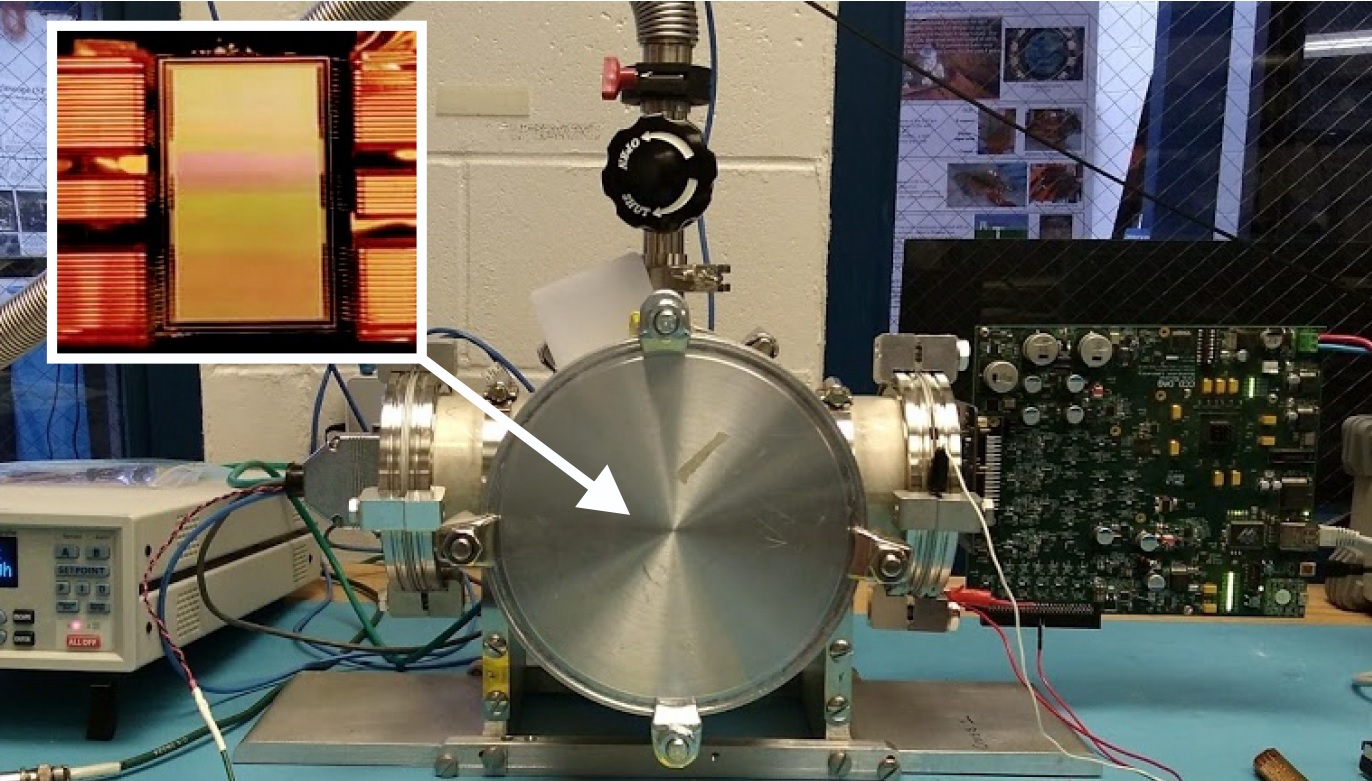}
        \caption{Skipper-CCD setup with LTA electronics.}
        \label{fig:experimental setup}
\end{center}
\end{figure}

\begin{figure}
\begin{center}
\includegraphics[width=\textwidth]{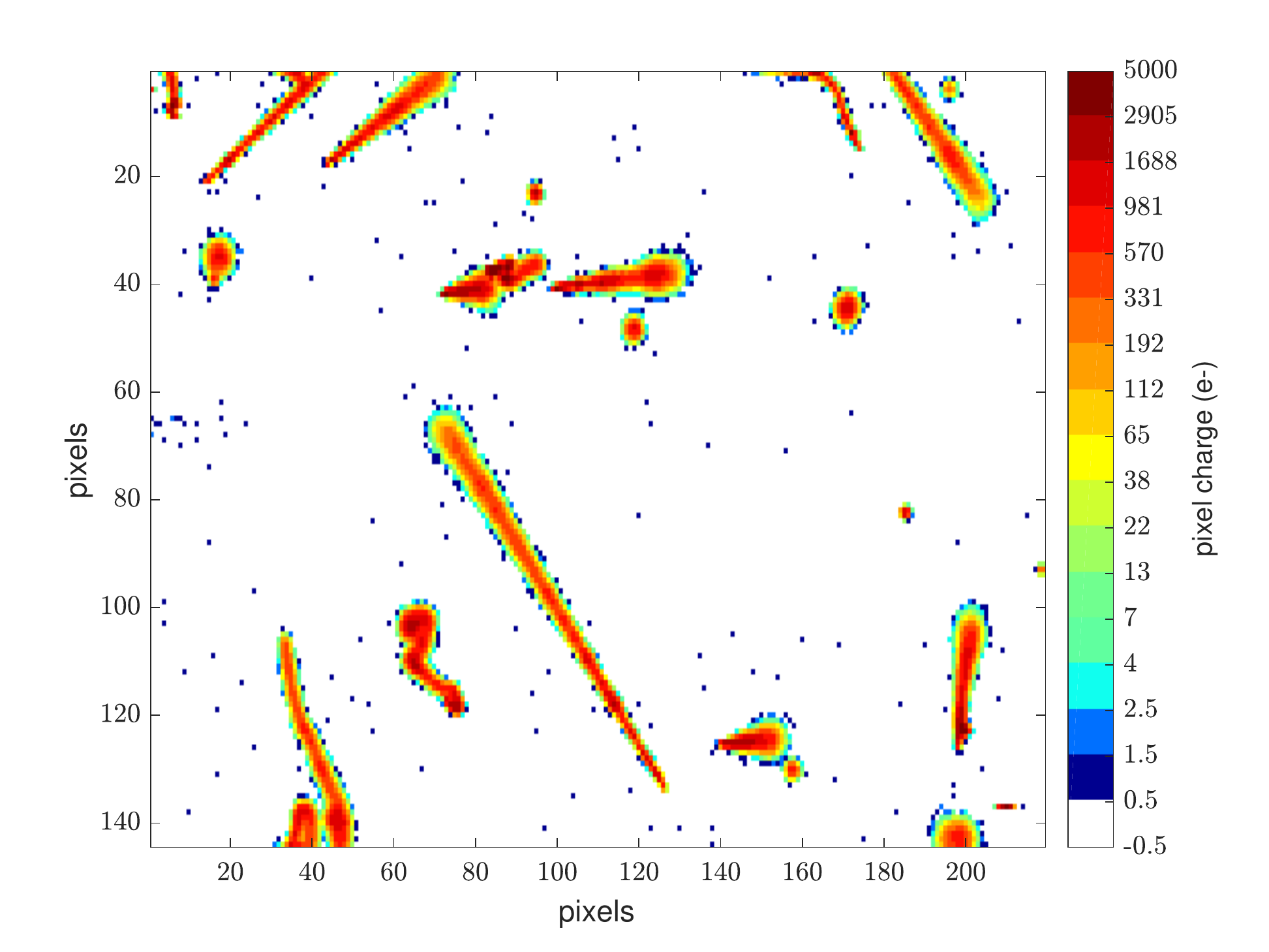}
        \caption{Small portion of an output image when reading the sensor with deep subelectron noise level(0.15e$^-$).}
        \label{fig:sample image}
\end{center}
\end{figure}

The experimental setup with an Skipper-CCD is shown in Figure \ref{fig:experimental setup}. The Skipper-CCD consist of silicon with a resistivity of 18~k$\Omega$-cm, an active area of 9.216~cm $\times$ 1.329~cm, a thickness of $675~\mu \text{m}$. The sensors were design by Steve Holland in Microsystems Laboratory in Lawrence Berkeley National Laboratory. No thinning process was applied to the back side. The CCD has four identical amplifiers, one in each corner. Each amplifier can read the entire CCD, but the usual mode of operation will be to read one quarter of the CCD consisting of 3072~rows and 443~columns of pixels. Each pixel has an area of $15~\mu \text{m} \times 15~\mu\text{m}$. A silicon-aluminum pitch adapter and copper-Kapton flex cable were glued and wirebonded to the Skipper-CCD. The CCD was operated at a temperature of 135~K. The cooling system is based on a closed circuit Helium based cryocooler \cite{cryomech}. A cold finger penetrates the dewar and connects to a copper mass on which the CCD sits flat. A temperature control system \cite{temp_controller} with a 50 Watt heater is used to keep the temperature of the CCD constant. Previously to cooling the dewar achieves vacuum (e.g. $10^{-5} torr$ ) using a \cite{pfeiffer} turbo-pump. The flex cable ends in a DB50 connector that connects to the interface electronics. The interface electronics connects directly to the 50 pin vacuum feed-through connector. The LTA connects to the outside part of the 50 pin connector. The vacuum and cooling lines are electrically isolated from the CCD system signal ground to avoid injecting low frequency noise into the CCD readout system. The LTA powers from +12V DC (upper right corner of Figure \ref{fig:experimental setup}) and connects to the PC through an Ethernet cable (bottom right corner of Figure \ref{fig:experimental setup}).
The DAQ software runs on the PC and generates images in FITS file format. The DAQ can also control parameters (e.g. voltages, timing) and get status through the on-board telemetry. 
Figure \ref{fig:sample image} shows a portion of one of the images taken with the system. The CCD was expose 6 hours in a light tight environment.The color scale shows the charge of each pixel. The color bins do not always represent the same charge packet values. The first colors are exaggerated to clear visualize pixels with less than four electrons. The figure shows that there is no mistake to identify pixels with no charge from those with charge. Single electrons in single pixels represent the dark current contribution of charge in the sensor.  The long traces are mostly atmospheric muons crossing the detector. One X-ray event is also observable. The good energy resolution can be also observed in the edges of these high energy events where only just a few electrons are observed. 

\subsection{Gain calibration}

\begin{figure}
\includegraphics[width=1\textwidth]{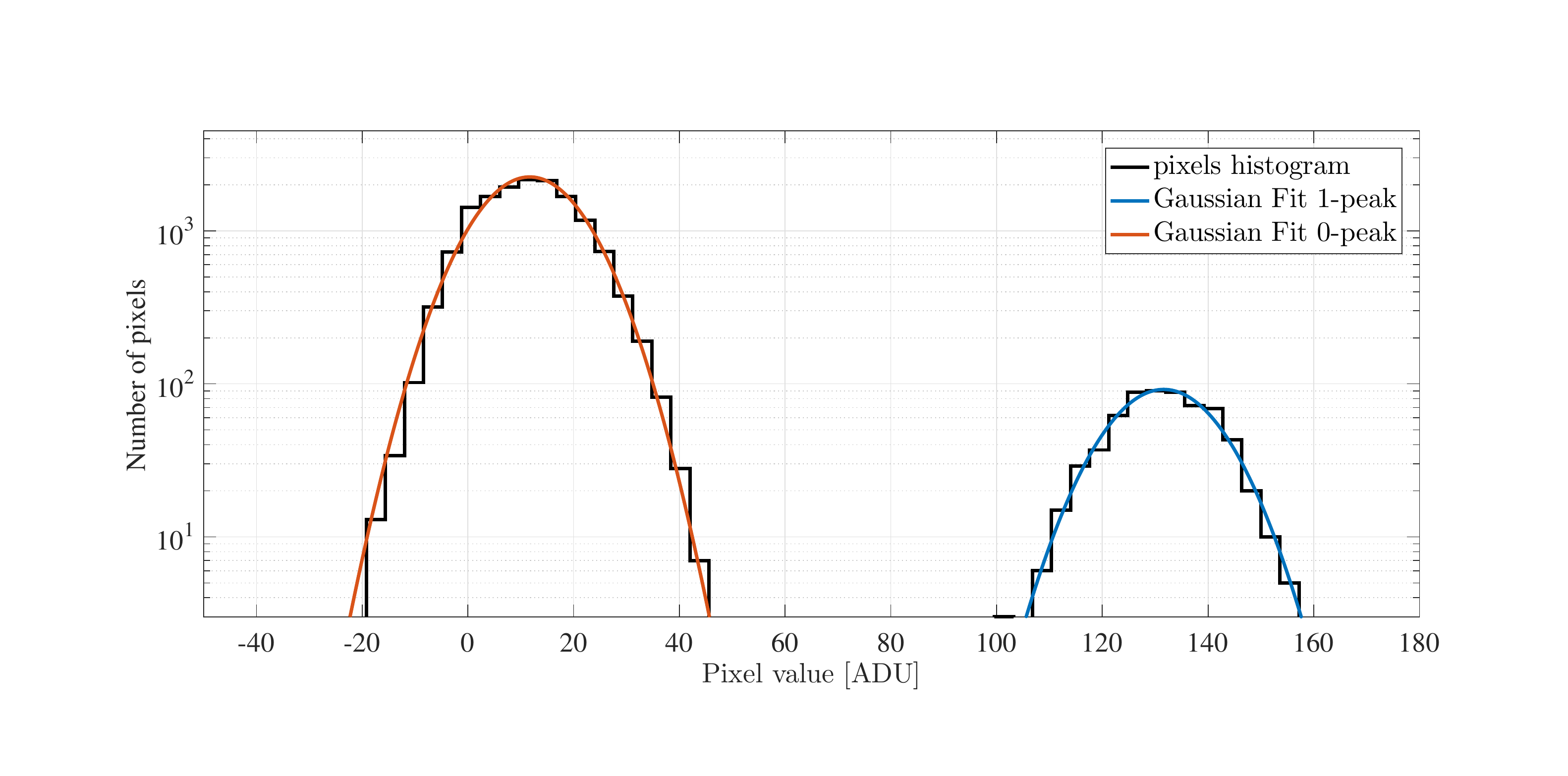}
        \caption{Pixel histogram showing the zero and one electrons peak for an integration time of $t_i=8~\mu$s. The solid lines show the best fit to each of the peaks. The pixel values is shown in ADU (analog-to-digital-converter units). The difference between the mean of each of the fits is used to calibrate the system and obtain the gain in ADU/e$^{-}$.}
\label{fig:gain histogram}
\end{figure}

The single charge counting capability of the Skipper-CCD was used to measured the gain of the video stage of each quadrant of the CCD. As it was explained the pixels are measured with deep subelectron noise so that the charge of each pixel can be determined unequivocally. Figure \ref{fig:gain histogram} shows the histogram for pixels with zero and one electron collected. Well defined peaks are observed due to charge quantization. The measure of the difference of the mean between two consecutive peaks gives the relation between ADUs and electrons of the entire signal video chain. The best normal fit to the peaks are used to estimate the gain as the difference of the fitted mean values.

\begin{figure}

\includegraphics[width=1\textwidth]{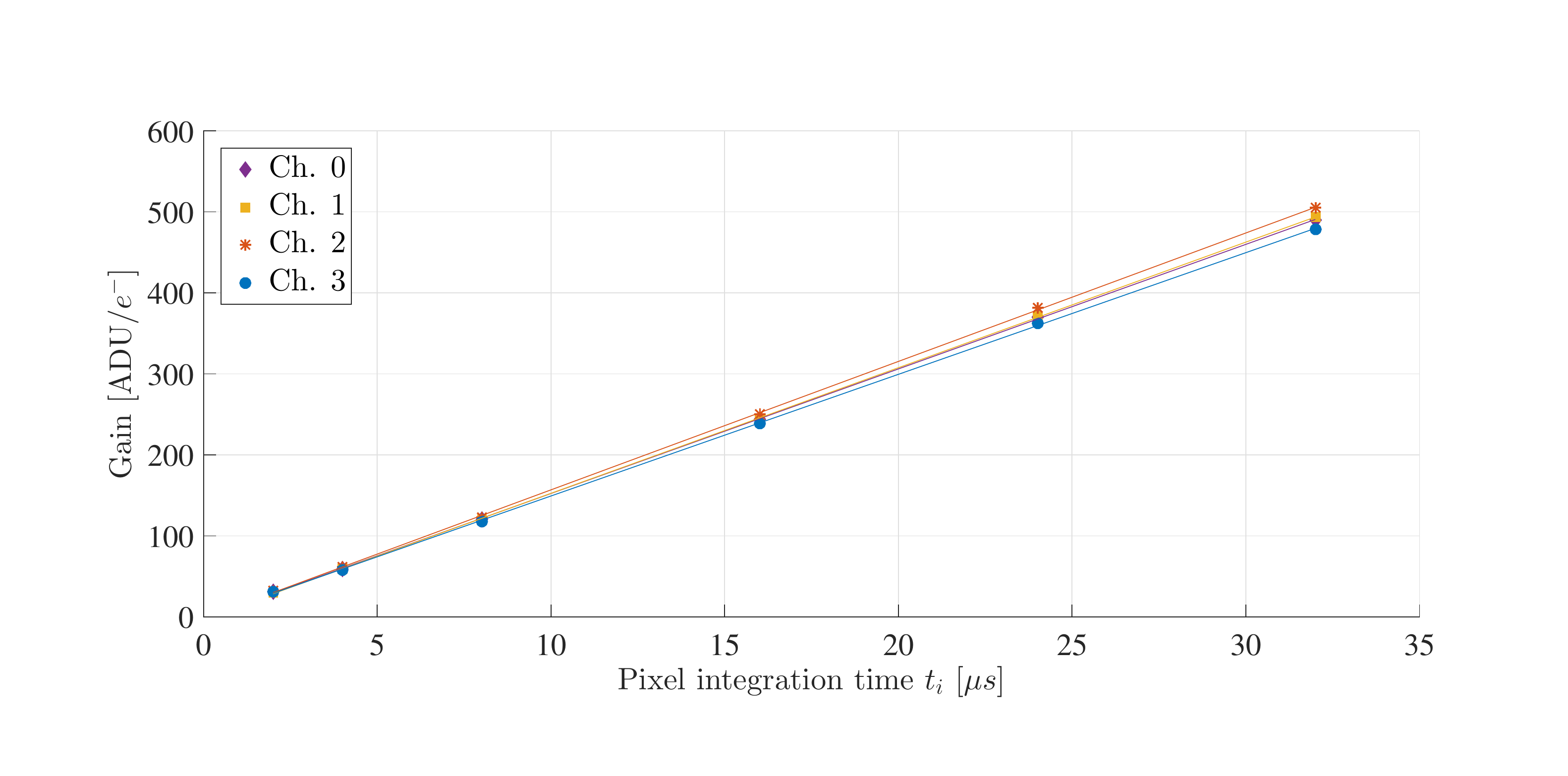}
        \caption{Markers show the measured channel gain as a function of the pixel integration time $t_i$. The minimum square error fit for each channel are also shown revealing the linearity of the gain with $t_i$.}
\label{fig:gain}
\end{figure}

The same procedure was repeated for the four amplifiers. The obtained gains of the system are shown in Fig. \ref{fig:gain} as a function of the integration time ($t_i$) of the individual measurement of each pixel. We define the integration time $t_i$ as the time interval during which the samples of the video signal are averaged to obtain either the pedestal or signal level. The solid lines in Fig. \ref{fig:gain} are minimum-square-error fits of a straight line to the points, which shows a linear and predictable response of the system. 

\subsection{Integration time scan}
\label{Sec:NoiseScan}
\begin{figure}
    \centering
    \includegraphics[width=1\textwidth]{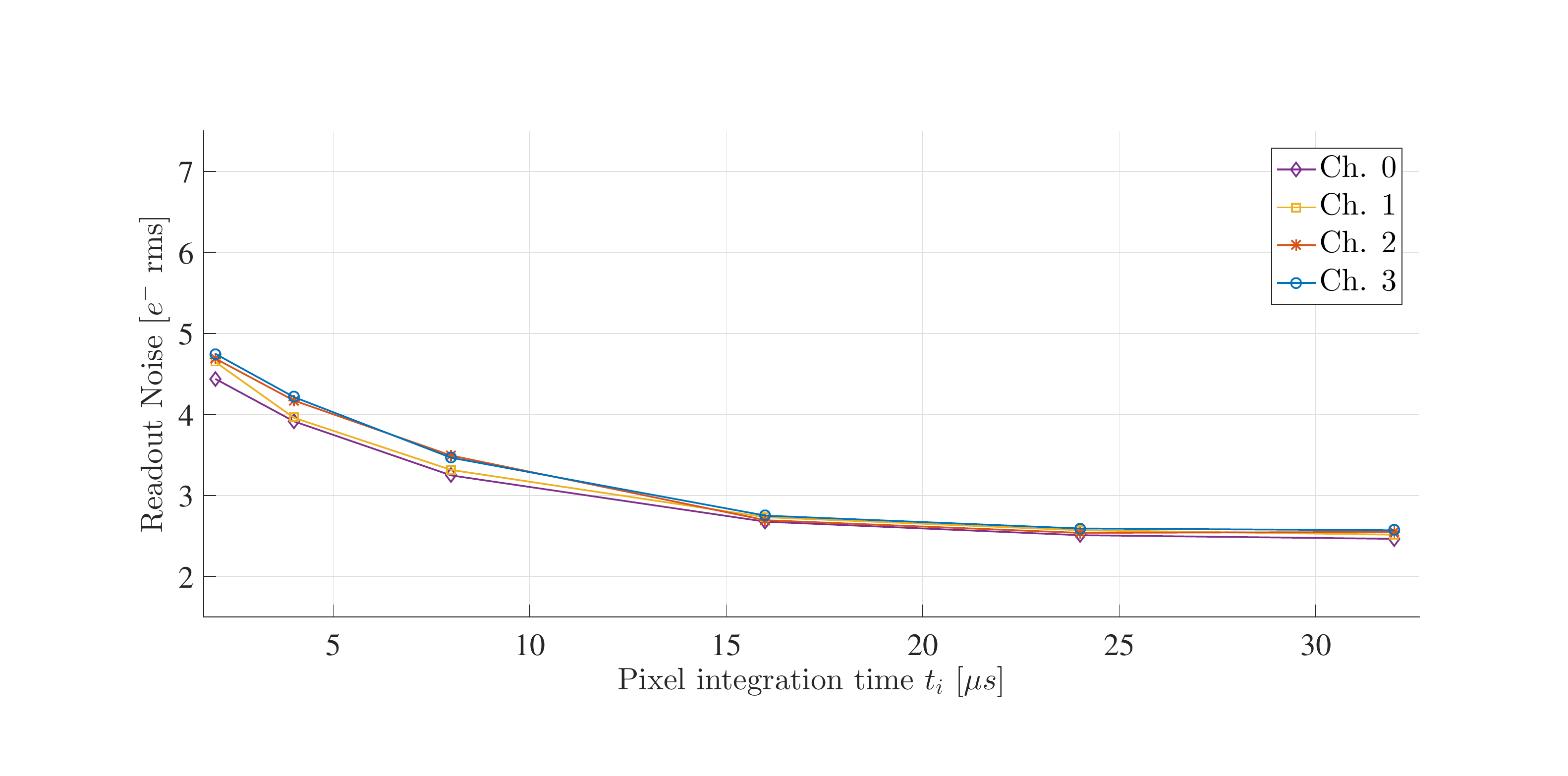}
    \caption{Noise scan measurement. Noise of the setup measured in electrons as a function of the pixel integration time $t_i$. The integration time $t_i$ is defined as the time during which the samples of the video signal are averaged to obtain either the pedestal or signal level.}
    \label{fig:noiseSCAN}
\end{figure}

The readout noise as a function of the integration time $t_i$ is a typical performance test of any CCD camera system. For this experiment only one skipper sample $N=1$ is used. Markers in Figure \ref{fig:noiseSCAN} shows the result of the noise scan measurements using the four channels. As expected, the noise is reduced when $t_i$ is increased achieving around $2.5$ electrons of noise for the higher values.

\subsection{Skipper scan and Gaussianity of the readout noise}

In this section we present the experimental results of the readout noise for a sweep of the number of measurements $N$ taken for each pixel. Given that the capability of taking multiple, non-destructive, independent samples of the same pixel charge is possible using the Skipper-CCD, we call this experiment Skipper scan. Each sweep is computed for a fixed pixel integration time $t_i$.

The ability of the system for acquiring images that contain subregions with different number $N$ of samples per pixel is used for this measurement. Images with seven different regions with $N=1, 10, 50, 100, 500, 1000, 5000$ samples are used. Each subregion for a fixed $N$ has around $650$ pixels, several images are acquired so that tens of thousands of pixels are used to compute each measurement of noise, most of those pixels are empty pixels.

To compute the Skipper scan a basic and standard preprocessing of the images for astronomy is performed for each of the subregions corresponding to a fixed value of $N$. First the mean is removed: if $P_i^j$ is the original value of the $i$-th pixel from the $j$-th image then
\begin{equation}
    \hat{P}_i^j=P_i^j-\bar{P}^j,
\end{equation}
is the value of each pixel with the mean $\bar{P}^j$ of the subregion removed. The mean $\bar{P}^j$ is computed with all the pixels of the same image and subregion of $N$, discarding possible outliers values caused by unwanted charge in those pixels. The outliers are identified as pixel with more than three times the normalized Median Absolute Deviation (MAD).
In the second step the baseline of the images is removed, to perform this operation the median $M$ of the images is computed and subtracted to each of the images, if $M_i$ is the $i$-th pixel of the median then
\begin{equation}
    \tilde{P}_i^j =\hat{P}_i^j-M_i.
\end{equation}
is the $i$-th pixel of the $j$-th image with the mean and baseline removed. The  $\tilde{P}_i^j$ pixels of the same subregion of all the images are collected together and used to compute the noise of the system.  

\begin{figure}
    \centering
    \subfloat[]{\includegraphics[width=1\textwidth]{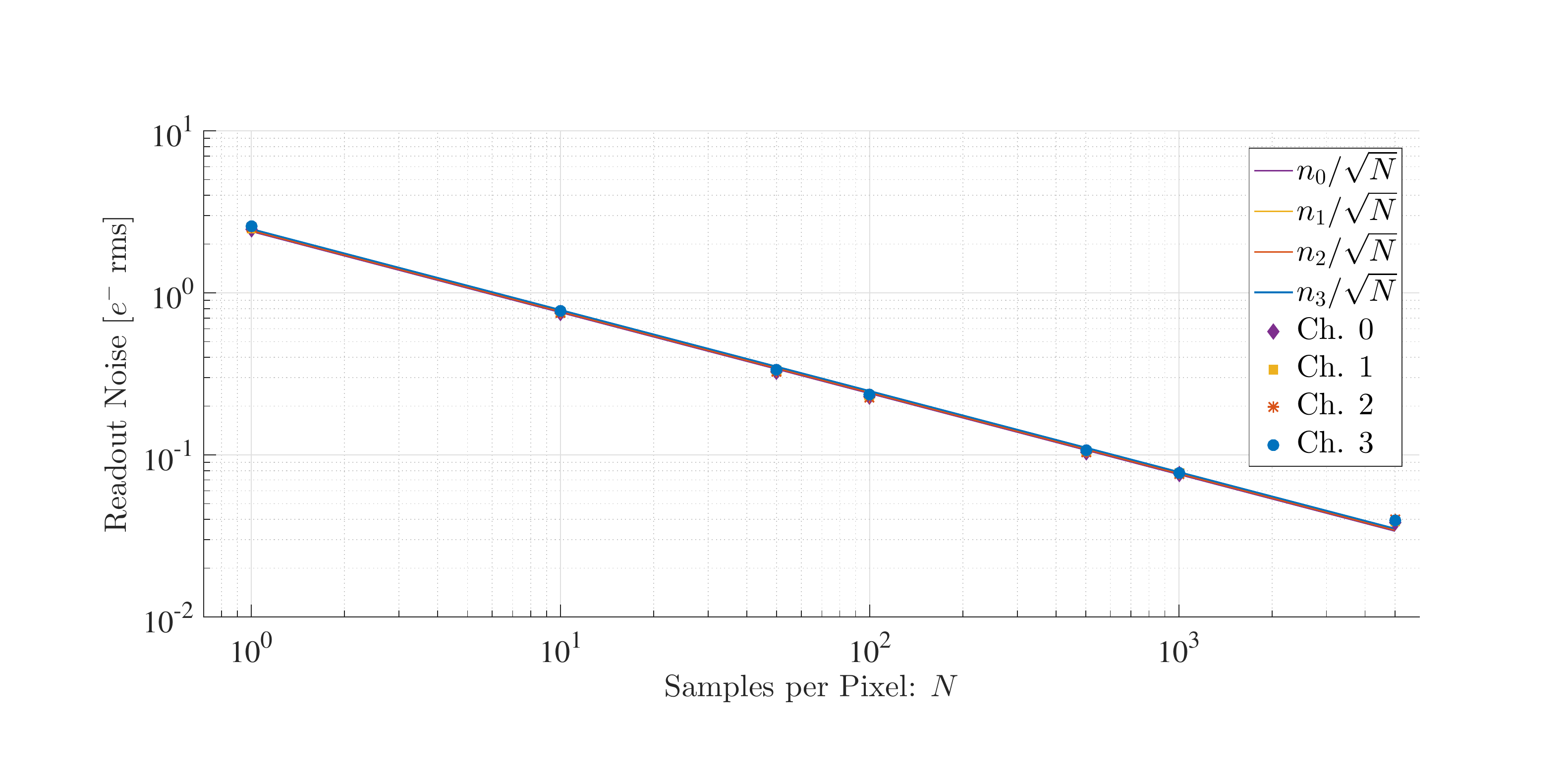}
    \label{Fig:skipperScan480}}\\
    \subfloat[]{\includegraphics[width=1\textwidth]{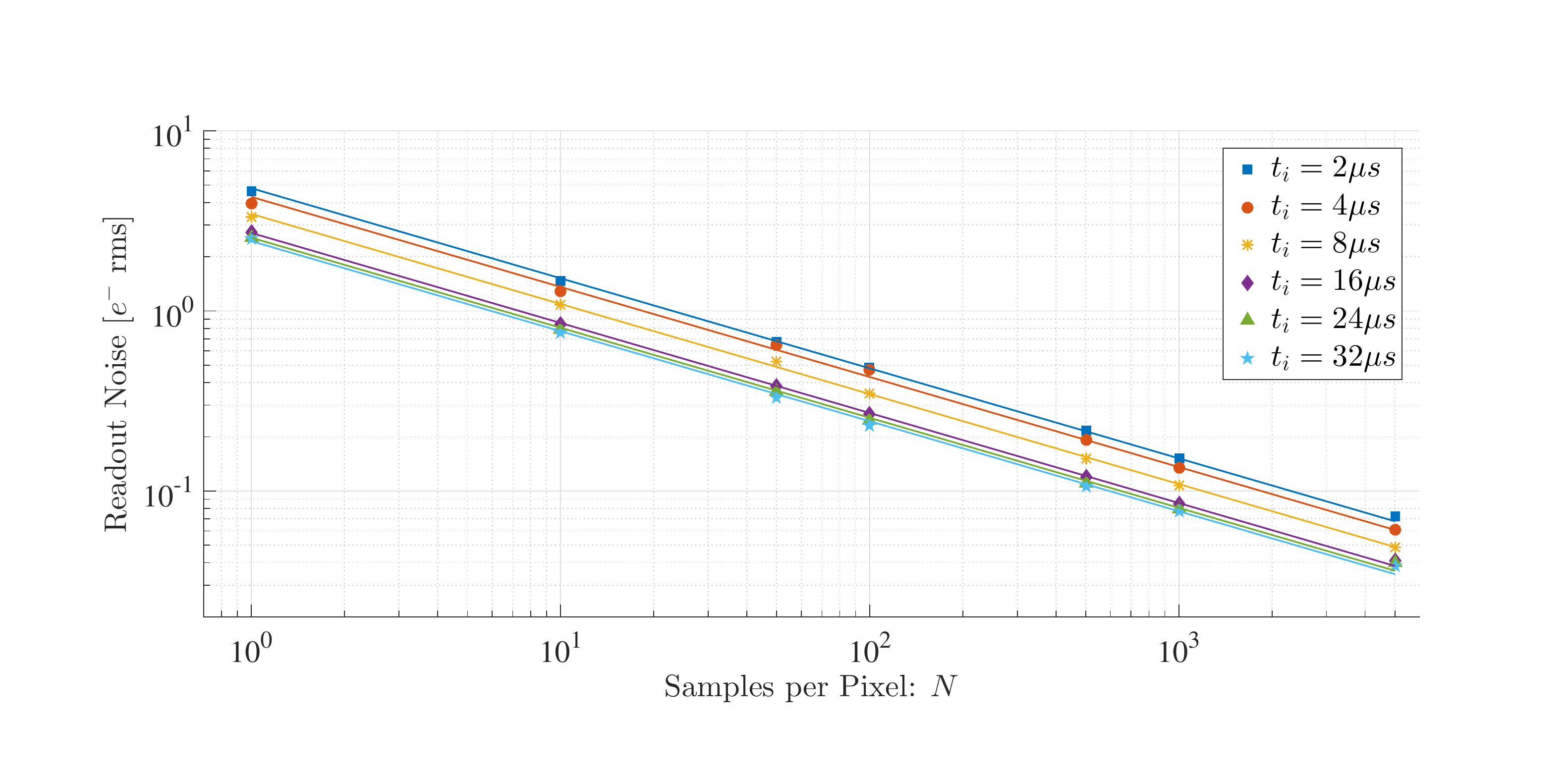}
    \label{Fig:skipperScanParamSsamp}}
    \caption{Skipper scan measurements: sweep of the readout noise as a function of the number of samples $N$ taken for each pixel. Figure \ref{Fig:skipperScan480} shows the Skipper scan for the four channels and an integration time of the pixel of $t_i=32$ $\mu $s; Figure \ref{Fig:skipperScanParamSsamp} show the Skipper scan for only one channel (Channel xx) parametrized for different integration times of the pixel: $t_i=2, 4, 8, 16, 24, 32$ $\mu s$. In both figures measurements are indicated with markers and the straight lines show the theoretical prediction $\propto 1/\sqrt{N}$.}
    \label{Fig:skipperScan}
\end{figure}

Figure \ref{Fig:skipperScan} shows the results of the Skipper scan test. The Skipper operation correspond to averaging $N$ independent noise sources (empty pixels) which results in a reduced readout noise. This reduction is theoretically proportional to $1/\sqrt{N}$ and becomes a straight line in a log-log plot which is also shown in  Fig. \ref{Fig:skipperScan} with solid lines. To estimate the proportionality constants $n_0 \dots n_3$ indicated in the legend of Fig. \ref{Fig:skipperScan480} least squares minimization using the points of each channel was applied. Figure \ref{Fig:skipperScan480} depicts the noise in electrons as a function of $N$ for the four channels available in the system and for an integration time of the pixel of $t_i=32$ $\mu s$. The experimental data is indicated with markers (diamonds, squares, asterisks and  solid circle), this plot reveals that all channels are almost superimposed, indicating that the noise performance of all the channels is the same, which is a desirable property. It can also be observed that the measurements follow the predicted rate of noise reduction as a function of $N$. The lower noise is achieved for $N=5000$ and is around $0.039$ $e^{-}$RMS.

Figure \ref{Fig:skipperScanParamSsamp} shows the results of skipper scan for channel 1 using different pixel integration times: $t_i=2, 4, 8, 16, 24, 32$ $\mu s$. Parallel lines following the theoretical $\propto 1/\sqrt{N}$ prediction  are observed in all cases. In agreement with results in Section \ref{Sec:NoiseScan}, the noise reduction (separation between lines) is more significant when increasing $t_i$ for the lower values, for example, when increasing the integration time from $t_i=4$ $\mu s$ to $t_i=8$ $\mu s$, and reaches a plateau for higher values of $t_i$, for example when increasing from $t_i=16$ $\mu s$ to $t_i=24$ $\mu s$ or $t_i=24$ $\mu s$ to $t_i=32$ $\mu s$. This is the expected behavior as the noise scan in Fig. \ref{fig:noiseSCAN} shows.

\begin{figure}
    \centering
    \includegraphics[width=1\textwidth]{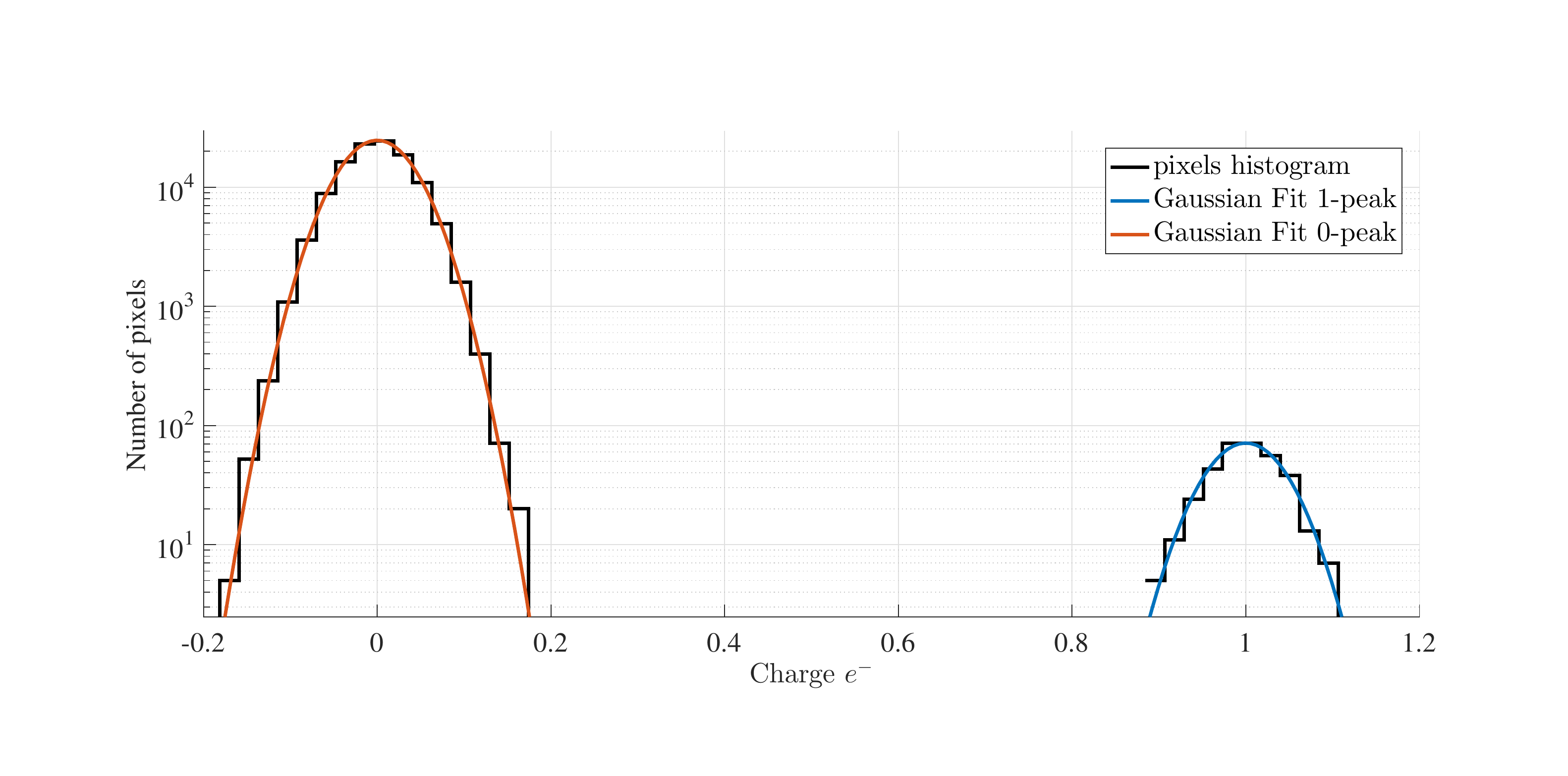}
    \caption{Histogram of the pixels with charge in the range between $0$ $e^{-}$ and 1$e^{-}$ and Gaussian FIT for peaks of $0$ and $1$ electron. The $y$-axis in logarithmic scale to show Gaussianity of the readout noise in four orders of magnitude. The histogram was computed for channel 2, an integration time of $t_i=32$ $\mu s$ and $N=5000$ samples for each pixel.}
    \label{fig:gaussianidad}
\end{figure}

Finally, Fig. \ref{fig:gaussianidad} shows the pixel histogram (in log scale) corresponding to the parameters $t_i=32$ $\mu s$ and $N=5000$. The peaks at $0$ electrons and $1$ electrons were obtained by fitting a Normal distribution to the data. The estimated standard deviation is the square root of the unbiased estimate of the variance. The logarithmic scale reveals that the data follows the normal fit more than four orders of magnitudes showing the gaussianity of the readout noise in a wide range which is also a desirable property.


\section{Conclusion}
\label{sec:conclusions}

In this work the first CCD controller specially designed for Skipper-CCD was presented. This type of CCD allows to readout nondestructively the same pixel charge multiple times, which reduces the readout noise. To drive skipper-type CCDs it is necessary to have the flexibility of sequencing the clocks in a different way compared to standard CCDs. The controller is a fully digital, single board, 4-channels, Ethernet based system which provides the necessary signals and sequencing flexibility. A detailed description of the main components and functionalities of the system were presented. The results from standard tests for CCD were detailed to show its performance. 



\bibliographystyle{IEEEtran}
\bibliography{main.bib}

\end{document}